\documentclass[a4paper,fleqn,usenatbib]{mnras}

\usepackage{newtxtext,newtxmath}

\usepackage[T1]{fontenc}
\usepackage{ae,aecompl}

\usepackage{graphicx}	
\usepackage{amsmath}	

\usepackage{adjustbox} 
\usepackage{lipsum}
\usepackage{caption} 
\usepackage[para,online,flushleft]{threeparttable} 

\usepackage{booktabs} 

\usepackage{url} 

\newcommand*\diff{\mathop{}\!\mathrm{d}}

\title[Vertical evolution of exocometary gas I]{Vertical evolution of exocometary gas: I. How vertical diffusion shortens the CO lifetime }
\author[S. Marino]{ S. Marino$^{1,2}$\thanks{E-mail:
    sebastian.marino.estay@gmail.com}, G. Cataldi$^{3,4}$, M.
  R. Jankovic$^{2}$, L. Matr\`a$^{5}$, and M. C. Wyatt$^{2}$ \\
  $^{1}$Jesus College, University of Cambridge, Jesus Lane, Cambridge CB5 8BL, UK\\
  $^{2}$Institute of Astronomy, University of Cambridge, Madingley Road, Cambridge CB3 0HA, UK\\
  $^{3}$Department of Astronomy, Graduate School of Science, The University of Tokyo, Tokyo 113-0033, Japan\\
  $^{4}$National Astronomical Observatory of Japan, Osawa 2-21-1, Mitaka, Tokyo 181-8588, Japan\\
  $^{5}$School of Physics, Trinity College Dublin, the University of Dublin, College Green, Dublin 2, Ireland}
\date{Accepted XXX. Received YYY; in original form ZZZ}

\pubyear{2022}

\begin{document}
\label{firstpage}
\pagerange{\pageref{firstpage}--\pageref{lastpage}}
\maketitle


\begin{abstract}

 
 Bright debris discs can contain large amounts of CO gas. This gas was
 thought to be a protoplanetary remnant until it was recently shown
 that it could be released in collisions of volatile-rich solids. As
 CO is released, interstellar UV radiation photodissociates CO
 producing CI, which can shield CO allowing a large CO mass to
 accumulate. However, this picture was challenged because CI is
 inefficient at shielding if CO and CI are vertically mixed. Here, we
 study for the first time the vertical evolution of gas to determine
 how vertical mixing affects the efficiency of shielding by CI. We
 present a 1D model that accounts for gas release, photodissociation,
 ionisation, viscous evolution, and vertical mixing due to turbulent
 diffusion. We find that if the gas surface density is high and the
 vertical diffusion weak ($\alpha_{\rm v}/\alpha<[H/r]^2$) CO
 photodissociates high above the midplane, forming an optically thick
 CI layer that shields the CO underneath. Conversely, if diffusion is
 strong ($\alpha_{\rm v}/\alpha>[H/r]^2$) CI and CO become well mixed,
 shortening the CO lifetime. Moreover, diffusion could also limit the
 amount of dust settling. High-resolution ALMA observations could
 resolve the vertical distribution of CO and CI, and thus constrain
 vertical mixing and the efficiency of CI shielding. We also find that
 the CO and CI scale heights may not be good probes of the mean
 molecular weight, and thus composition, of the gas. Finally, we show
 that if mixing is strong the CO lifetime might not be long enough for
 CO to spread interior to the planetesimal belt where gas is produced.
 
\end{abstract}

\begin{keywords}
  circumstellar matter - planetary systems - methods: numerical
\end{keywords}



\section{Introduction}
\label{sec:intro}





Over the last decade it has become increasingly clear that
circumstellar gas is present beyond the short ($\lesssim10$~Myr)
protoplanetary disc phase in exoplanetary systems. This gas was first
found decades ago \citep{Slettebak1975, Zuckerman1995}, but its
ubiquity in systems with bright debris discs was only recently
revealed \citep[e.g.][]{Iglesias2018, Rebollido2020, Moor2017}. These
dusty discs were known to be the result of collisions of large (at
least km-sized) planetesimals \citep{Wyatt2008, Hughes2018,
  Marino2022}, but for a long time were considered gas-free. Today,
gas has been discovered in tens of systems mainly through CO emission
lines around AFGM-type stars tracing cold gas \citep[e.g.][]{Dent2014,
  Marino2016, Kral2020COsurvey, Matra2019twa7}, and UV and optical
absorption lines tracing hot gas mainly around BA-type stars
\citep{Montgomery2012, Iglesias2018, Rebollido2018}. The presence of
gas for hundreds or thousands of Myr could have strong implications
for the atmospheres of exoplanets as these could accrete this gas
delivering volatiles to their atmospheres
\citep{Kral2020atmospheres}. Moreover, gas could also shape the dust
spatial distribution \citep{Takeuchi2001, Thebault2005, Krivov2009,
  Lyra2013photoelectric, Pearce2020, Olofsson2022}, and thus
understanding its evolution is important to correctly interpret the
dynamics of these systems. In addition, in a few systems gas
absorption lines have been shown to be variable and Doppler shifted,
indicating that the hot gas originates from sublimating exocomets on
highly eccentric orbits \citep[e.g.][]{Kennedy2018febs}. In this
paper, we focus on the study of cold gas found at tens of au.

Where this cold gas originates from and how it evolves are one of the
key current questions in debris disc studies. Two main scenarios have
been proposed to explain it. First, the gas could be a leftover from
the protoplanetary disc phase and dominated by hydrogen. In this
scenario, some debris discs could be hybrid containing primordial gas
that has not dispersed yet and secondary dust created by collisions
\citep{Kospal2013}. The hybrid scenario is motivated by the fact that
a considerable fraction of debris discs around young A-type stars are
gas-rich, containing vast amounts of CO gas that are comparable to
protoplanetary discs around Herbig AeBe stars
\citep[][]{Moor2017}. Recent simulations have shown that primordial
gas could survive for longer than expected in optically thin discs
around A-type stars due to a weaker photoevaporation compared to
later-type stars \citep{Nakatani2021}. This could potentially explain
why gas-rich discs are found around A-type stars preferentially. This
model, however, has not yet been tested against the measured CO gas
masses and constraints on its composition. As a way to determine the
gas origin, recent surveys have searched for multiple molecules (apart
from CO) that are normally detected in protoplanetary discs
\citep{Klusmeyer2021, Smirnov2021}. Those studies, however, have only
led to non-detections implying abundances relative to CO orders of
magnitude lower than those in protoplanetary discs. These results do
not rule-out that the gas is primordial since, as shown by
\cite{Smirnov2021}, the optically thin nature of debris discs means
most molecules will be short-lived and scarce.

On the other hand, the gas could be of secondary origin, i.e. released
from volatile-rich solids as they fragment and grind down in the
collisional cascade \citep{Zuckerman2012, Dent2014}. In this scenario,
hydrogen-poor gas is continuously released and could viscously evolve
spreading beyond the planetesimal belt \citep{Kral2016}, or could be
blown out via stellar winds or radiation pressure if densities are low
or the star has a high luminosity \citep{Kral2021,
  Youngblood2021}. This exocometary origin scenario has successfully
explained the low levels of CO gas found in a few systems
\citep[e.g.][]{Dent2014, Marino2016, Matra2017fomalhaut}. Once the CO
gas is released, it should only survive for ${\sim}130$~yr due to
interstellar UV photodissociating radiation \citep{Visser2009,
  Heays2017}, and thus the observed amount has been used to infer the
rate at which CO gas was released. Since the gas release requires
collisions between solids to free trapped gas or expose ices from
their interiors, the gas release rate is a fraction of the rate at
which mass is lost in the cascade. This fraction is approximately the
mass fraction of volatiles in the planetesimals feeding the cascade,
and was found to be consistent with Solar System comets
\citep{Marino2016, Matra2017fomalhaut}. If these planetesimals are
similar in composition to Solar System comets, we would expect other
molecules to be released as well. However, their short lifetime and
low abundance in comets relative to CO makes their detection very
challenging \citep{Matra2018}.

The main problem with the secondary origin scenario has been to
explain the several A-type stars with vast amounts of CO gas
\citep[e.g.][]{Moor2017}, which suggest CO lifetimes are much longer
or gas release much higher than expected. \cite{Kral2019} found a
solution by proposing that it is the carbon atoms, from previously
photodissociated CO, that act as a shield. Carbon in neutral form (CI)
has an ionisation cross section that covers all the CO
photodissociation bands, and thus could become an effective shield
prolonging the CO lifetime \citep{Rollins2012}. In fact, CI has been
found in some of these gas-rich/shielded discs adding supporting
evidence to this scenario \citep{Kral2019, Moor2019,
  Higuchi2019}. This model is also successful at explaining the
amounts of gas in the population of A-type and FGK-type stars with
debris discs \citep{Marino2020gas}.


This solution and success of the secondary origin scenario to explain
even the CO-rich systems has been recently
challenged. \cite{Cataldi2020} pointed out that carbon is only an
effective shield if it is vertically distributed in a geometrically
thin, but optically thick surface layer completely far above and below
the CO-rich midplane. Instead, if carbon and CO are vertically mixed,
carbon shielding is weak and the observed vertical column densities
may not be enough to explain the amount of CO gas in these discs. This
finding emphasises the importance of the vertical processes, in
addition to the radial ones, in the evolution of the gas. Previous
theoretical studies on exocometary gas have only looked at the radial
evolution \citep[e.g.][]{Kral2016, Marino2020gas}. This is why in this
paper we focus for the first time in trying to understand the vertical
evolution of the gas and how vertical diffusion affects carbon
shielding. In \S\ref{sec:model} we present and describe a new model
that takes into account photodissociation, vertical diffusion,
ionisation, and viscous evolution of gas continuously released in a
planetesimal belt. \S\ref{sec:results} presents the main findings and
evolution of the gas for different model parameters. We discuss these
results in \S\ref{sec:discussion} in terms of its implications for the
known gas-rich debris discs, its uncertainties, how observations could
shed light on the vertical mixing of carbon and CO, and some of the
limitations of our model. Finally, in \S\ref{sec:conclusions} we
summarise the main conclusions of this paper.




\section{Model description}

\label{sec:model}

In this work, we focus on the vertical evolution of the exocometary
gas taking into account the release of CO gas, its photodissociation,
ionisation of CI, viscous evolution, and vertical mixing through
turbulent diffusion. This evolution is simulated with \textsc{exogas},
a \textsc{python} package that we have developed and made available at
\url{https://github.com/SebaMarino/exogas}. \textsc{exogas} consists
of two main modules, one to model the radial evolution of gas based on
the simulations developed in \citet{Marino2020gas}, and a new one that
focuses on the vertical evolution that is presented below.

The vertical model consists of a one-dimensional distribution of gas
as a function of height $z$ in the middle of a planetesimal belt of
radius $r$, where gas is being released. We assume that the released
gas is pure CO and thus the C/O ratio of the gas is equal to one
throughout our simulations. Note that the C/O ratio could be different
in reality if other gas species that are abundant in comets were
released as well (e.g. CO$_2$, H$_2$O). The presence of these
additional species is, however, very uncertain due to the lack of
observational constraints and the uncertain mechanism through which
gas is released in the cold outer regions. Therefore, the gas in our
model is composed of CO, its photodissociation products CI and OI, and
ionised species CII and electrons. Note that since the C/O ratio is
equal to one and we do not expect oxygen to be ionised
\citep{Kral2016}, the number density of oxygen is simply equal to the
carbon (CI+CII) number density and the number density of electrons is
equal to the number density of CII. Therefore, of these five species
only CO, CI, and CII need to be explicitly modelled and the total gas
density ($\rho$) is equal to $\rho_{\rm CO}+\frac{28}{12}(\rho_{\rm
  CI}+\rho_{\rm CII})$.




The evolution of these three species is summarised in Equations
\ref{eq:CO}, \ref{eq:CI} and \ref{eq:CII}. The CO gas evolution is
ruled by the release of new CO gas ($\dot{\rho}^{+}$), the
photodissociation of CO ($\dot{\rho}_{\rm ph}$), viscous evolution in
the radial direction ($\dot{\rho}_{\rm vis, CO}$), and vertical
diffusion ($\dot{\rho}_{\mathrm{CO,} D}$). The CI gas evolution is
ruled by the CO photodissociation, viscous evolution, diffusion, and
ionisation/recombination ($\dot{\rho}_{\rm ion}$). Finally, the
evolution of CII is ruled by viscous evolution, diffusion, and
ionisation/recombination.
\begin{align}
  & \frac{\partial \rho_{\rm CO}}{\partial t} = \dot{\rho}^{+}-\dot{\rho}_{\rm ph} - \dot{\rho}_{\rm vis, CO} +\dot{\rho}_{\mathrm{CO,} D}, \label{eq:CO}\\
  & \frac{\partial \rho_{\rm CI}}{\partial t} = \frac{12}{28}\dot{\rho}_{\rm ph} - \dot{\rho}_{\rm vis, CI} - \dot{\rho}_{\rm ion} +\dot{\rho}_{\mathrm{CI,} D}, \label{eq:CI} \\
  & \frac{\partial \rho_{\rm CII}}{\partial t} =  \dot{\rho}_{\rm ion} - \dot{\rho}_{\rm vis, CII} +\dot{\rho}_{\mathrm{CII,} D}. \label{eq:CII}
\end{align}
These equations are solved numerically using finite differences and
each individual processes (CO gas release, photodissociation,
ionisation, viscous evolution and diffusion) is described
below. Table~\ref{tab:parameters} summarises the definition and units
of the most important parameters and variables.

\subsection{Gas release}

We assume the gas is released with a Gaussian vertical distribution
with a scale height equal to $H$, i.e. with the same vertical
distribution as in hydrostatic equilibrium. This assumption together
with our viscous evolution treatment (\S\ref{sec:visevol}) keeps the
total gas density in hydrostatic equilibrium throughout the simulation
and allows us to avoid having to model net vertical motions of the gas
by solving the Navier-Stokes equations of classical fluid
dynamics. Thus, CO gas is released in our 1D grid at a rate of
\begin{align}
  & \dot{\rho}_{+}(z) = \dot{\Sigma}_{+} \frac{e^{- \frac{z^2}{2H^{2}}}}{\sqrt{2\upi} H}, \label{eq:rhop}\\
  & H=c_{\rm s}/\Omega_{\rm K}, \label{eq:H}
\end{align}
where $\dot{\Sigma}_{+}$ is the gas release rate per unit surface at
the center of the belt, $H$ is the gas scale height, $\Omega_{\rm K}$
is the Keplerian frequency, and $c_{\rm s}$ is the isothermal sound
speed. The sound speed and temperature ($T$)
are set by
\begin{align}
  & c_{\rm s} = \sqrt{\frac{k_{\rm B}T}{\mu m_{\rm p}}}, \label{eq:cs}\\
  & T = 278.3\ {\rm K}\ \left(\frac{L_{\star}}{L_{\odot}}\right)^{1/4} \left(\frac{r}{\rm 1\ au}\right)^{-1/2}, \label{eq:T}
\end{align}
where for simplicity we have assumed a constant temperature as a
function of height and time, and equal to the midplane equilibrium
temperature for a blackbody. This temperature is simply set by the
stellar luminosity, $L_{\star}$, and belt radius, $r$. The mean
molecular weight in \ref{eq:cs}, $\mu$, is assumed to be equal to 14,
i.e. equivalent to a gas dominated by carbon and oxygen in equal
proportions. A different temperature or $\mu$ (e.g. 28 if the gas is
CO dominated) would change the scale height and the gas disc could
become thicker or thinner. Nevertheless, the evolution of the gas
surface density and its overall vertical distribution (relative to
$H$) is not very sensitive to the value of $H$ itself. This is because
the CO photodissociation depends mainly on the optical depths and
column densities rather than the volumetric densities, and the
vertical diffusion timescale is independent of $T$ and $\mu$. This is
discussed and demonstrated in \S\ref{sec:highT}.


In this paper, we consider a scenario where collisions between solids
are responsible for the release of gas and thus the gas input rate is
proportional to the mass loss rate of solids \citep[e.g. as discussed
  in][]{Zuckerman2012, Matra2015, Marino2016}. Since the collisional
rates are proportional to the disc mass or its density
\citep{Wyatt2007hotdust}, $\dot{\Sigma}_{+}$ is thus proportional to
the squared density of solids. We assume that the radial distribution
of solids follows a Gaussian distribution with a
full-width-half-maximum (FWHM), $\Delta r$, and the total gas release
rate is $\dot{M}_{+}$. Therefore, $\dot{\Sigma}_{+}$ can be defined as
\begin{equation}
  \dot{\Sigma}_{+} = \frac{\dot{M}_{+}\sqrt{2 \ln(2)}}{\upi^{3/2} r \Delta r}. \label{eq:sigmap}
\end{equation}

\subsection{CO photodissociation}
\label{sec:photodissociation}

We want to estimate the photodissociation rate of CO molecules at each
height in the disc. In order to do this, we first define the
unshielded CO photodissociation rate per molecule due to the
interstellar radiation field (ISRF) as
\begin{equation}
  R_{\rm ph,0}= \int \sigma_{\rm ph}(\lambda) \phi_{\lambda} \ \diff \lambda, \label{eq:Rph0} 
\end{equation}
where $\sigma_{\rm ph}(\lambda)$ is the photodissociation cross
section of CO \citep{Heays2017}\footnote{The cross section of CO was
downloaded from
\url{https://home.strw.leidenuniv.nl/~ewine/photo/cross_sections.html}. They
correspond to an excitation temperature of 100~K and a Doppler
broadening of 1~km~s$^{-1}$}, and $\phi_{\lambda}$ is the ISRF
\citep{Draine1978, vanDishoeck1982}\footnote{Downloaded from
\url{https://home.strw.leidenuniv.nl/~ewine/photo/radiation_fields.html}}. This
rate is equal to the inverse of the photodissociation timescale of
130~yr.

If the CO or CI column densities are high enough, CO molecules will
become partly or fully shielded from the UV photodissociating
radiation. In general, this type of disc is relatively flat with
vertical aspect ratios ($H/r$) $\ll1$, and thus radiation entering the
disc horizontally through the disc midplane will encounter much higher
column densities than radiation entering vertically. Therefore,
shielding will be a function of the direction considered and the
height above the midplane. In order to simplify the calculations of
the CO photodissociation rate, we assume a plane-parallel model,
i.e. all disc quantities are constant within a plane at height
$z$. This means that at a given height $z$, the column density (or
optical depth) away from that point will only vary as a function of
the polar angle, $\theta$, measured with respect to the vertical
direction. This is an approximation, since in reality the density will
also vary (at least) radially. Nevertheless, we expect this to be a
minor effect, especially at the belt central radius. Therefore, the CO
photodissociation rate per molecule is defined as
\begin{equation}
  R_{\rm ph}(z) = \frac{1}{4\upi} \int \int \sigma_{\rm ph}(\lambda) \phi_{\lambda} e^{-\tau(\lambda, \theta, z)}\ \diff \lambda \diff \Omega, \label{eq:Rph} 
\end{equation}
where the optical depth ($\tau$) is defined from a point at height $z$
outwards in a direction $\theta$ and at a wavelength $\lambda$. Note
that $\tau$ is the sum of the optical depth of all species
(e.g. H$_2$, H, dust, CO, CI) that can absorb the ISRF at the UV
wavelengths that cause CO photodissociation. These are only CO and CI
in our model, and thus $\tau$ is defined as
\begin{equation}
  \tau(\lambda, \theta, z) = \tau_{\rm CO}(\lambda, \theta, z) + \tau_{\rm CI}(\lambda, \theta, z).
\end{equation}
Using the plane-parallel approximation, the optical depths are simply
\begin{equation}
  \tau_i (\lambda, \theta,z)= N_i^{\pm}(z) \sigma_i(\lambda)/|\cos(\theta)|,  \label{eq:tau}
\end{equation}
where $N_i^{\pm}(z)$ is the vertical column density above ($+$) $z$ for $\theta<\upi/2$ or below ($-$) $z$ for $\theta>\upi/2$
\begin{align}
  & N_{i}^{+}(z) = \int_{z}^{\infty} n_i(z') \diff z', \\
  & N_{i}^{-}(z) = \int_{-\infty}^{z} n_i(z') \diff z'. 
\end{align}

\begin{figure}
  \centering \includegraphics[trim=0.0cm 0.0cm 0.0cm 0.0cm,
    clip=true, width=1.0\columnwidth]{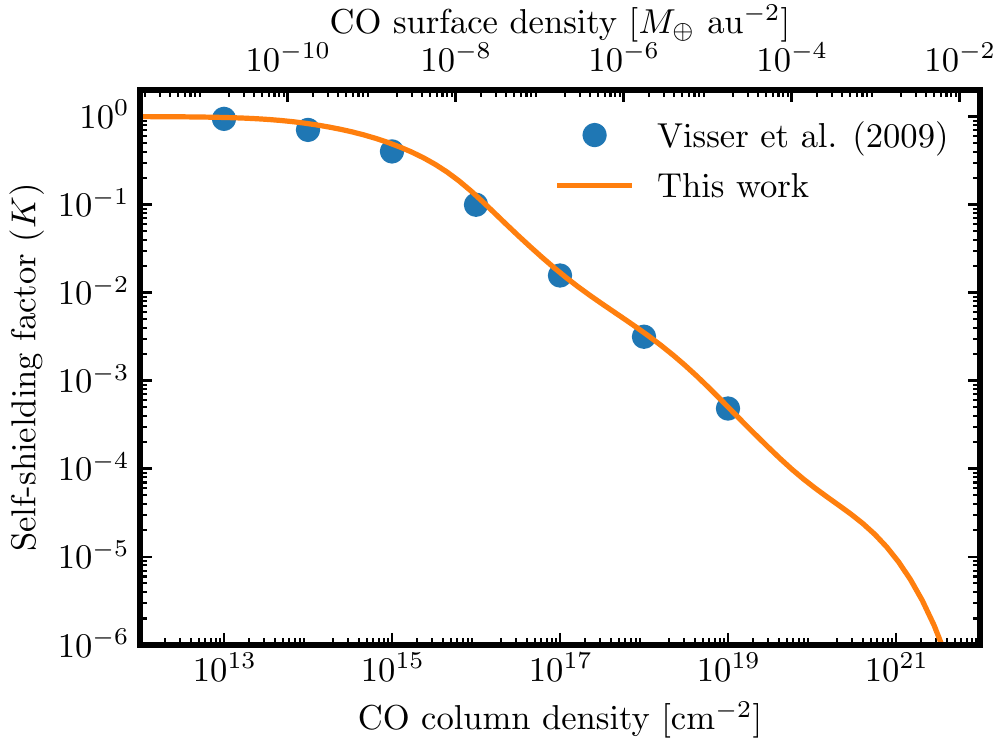}
  \caption{Self-shielding factor $K$ as a function of CO column
    density (orange). The blue points correspond to values in Table 6
    of \citet{Visser2009} for an excitation temperature of 50~K,
    Doppler broadening of 0.3 km~s$^{-1}$ and $^{12}$CO/$^{13}$CO
    abundance ratio of 69.}
  \label{fig:self-shielding}
\end{figure}

Since the CI cross section is roughly constant across the UV
wavelength range where CO photodissociation happens due to the ISRF
($90-110$~nm), we can write
\begin{equation}
  R_{\rm ph}(z)=\frac{1}{2} \int_{0}^{\upi}  \sin(\theta) e^{-\tau_{\rm CI}(\theta,z)}   \int_{\rm 90\ nm}^{\rm 110\ nm}  \sigma_{\rm ph}(\lambda) \phi_{\lambda} e^{-\tau_{\rm CO}(\lambda, \theta, z)}    \diff \theta \diff \lambda,
\end{equation}
where the solid angle integral was converted to an integral over the
polar angle only as nothing depends on the azimuthal angle
(plane-parallel approximation). To simplify this further, we define
the CO self-shielding factor $K$ as
\begin{equation}
  K(N^{\pm}_{\rm CO}(z)/|\cos(\theta)|) = \frac{\int \sigma_{\rm ph} \phi_{\lambda} e^{-\tau_{\rm CO}(\lambda, \theta, z)} \diff \lambda }{\int \sigma_{\rm ph} \phi_{\lambda} \diff \lambda }, \label{eq:K}
\end{equation}
which corresponds to the self-shielding function from
\citet{Visser2009}. The CO self-shielding factor can be understood as
the photodissociation rate (relative to the unshielded rate) for a CO
molecule if it was surrounded by a spherical cloud with a given column
density from its centre. Figure~\ref{fig:self-shielding} compares $K$
with the values from \citet{Visser2009} as a function of the CO column
density, showing a good agreement between the two. The self-shielding
factor is slightly higher (less shielding) at intermediate column
densities. This difference is likely due to the absence of CO
isotopologues $^{13}$C$^{16}$O and $^{12}$C$^{18}$O in our
calculations, and the use of an updated CO cross-sections table from
\cite{Heays2017}.

Finally, we can write the photodissociation rate per molecule and the
CO mass photodissociation rate per unit volume as
\begin{align}   
  & R_{\rm ph}(z) = \frac{R_{\rm ph,0}}{2} \int_{0}^{\upi} \sin(\theta) K\left(\frac{N^{\pm}_{\rm CO}}{|\cos(\theta)|}\right) e^{-N^{\pm}_{\rm CI}\sigma_{\rm CI}/|\cos(\theta)|}  \diff \theta \\
  & \dot{\rho}_{\rm ph}(z)=\rho_{\rm CO}(z) R_{\rm ph}(z). \label{eq:rhoph}
\end{align}

\begin{figure}
  \centering
  \includegraphics[trim=0.0cm 0.0cm 0.0cm 0.0cm, clip=true,
    width=1.0\columnwidth]{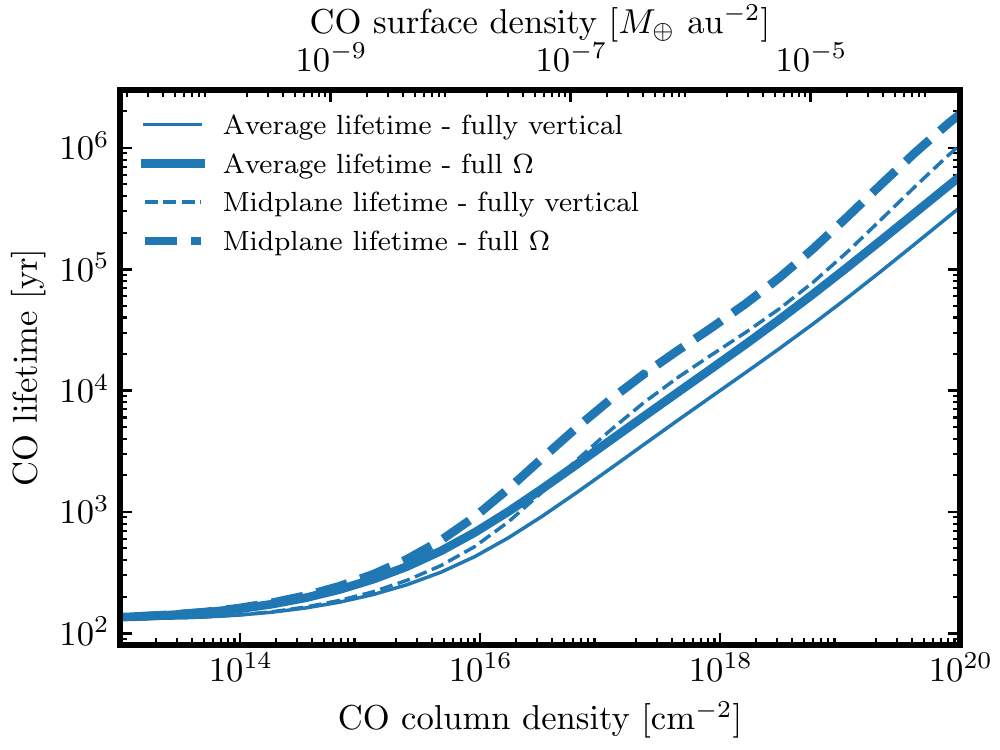}
  \caption{CO lifetime as a function of the CO vertical column density
    in the absence of CI. The solid lines show the average lifetime,
    while the dashed lines show the lifetime in the midplane. The thin
    and thick lines correspond to considering ISRF photons enter the
    disc only vertically or from all directions, respectively. }
  \label{fig:tCO_co}
\end{figure}

Figure~\ref{fig:tCO_co} shows the CO lifetime as a function of the CO
vertical column density in the absence of CI. The solid lines
correspond to the average lifetime defined as $\tau_{\rm CO, mean} =
\int \rho_{\rm CO}(z) \diff z /\int \dot{\rho}_{\rm ph}(z) \diff z$,
when considering the ISRF photons to enter the disc only vertically
(thin line) or to be distributed in all directions (thick line). We
find that the CO lifetime can be underestimated by a factor
${\sim}1.8$ for column densities above $10^{16}$~cm$^{-2}$ when the
ISRF is assumed to enter the disc only vertically (where the optical
depth is lowest) as done in \citet{Kral2019, Marino2020gas,
  Cataldi2020}. The dashed lines show the CO lifetime at the
midplane. These show that the midplane lifetime is longer by a factor
$1.5-3$ for column densities above $10^{16}$~cm$^{-2}$. Thus
considering the midplane lifetime as representative \citep[as done
  in][]{Kral2019, Marino2020gas} would overestimate the true CO
lifetime in the disc. Nevertheless, both effects balance out and the
average lifetime considering all directions of radiation is very close
to the midplane lifetime when assuming all UV radiation enters the
disc vertically.

\begin{figure}
  \centering \includegraphics[trim=0.0cm 0.0cm 0.0cm 0.0cm, clip=true,
    width=1.0\columnwidth]{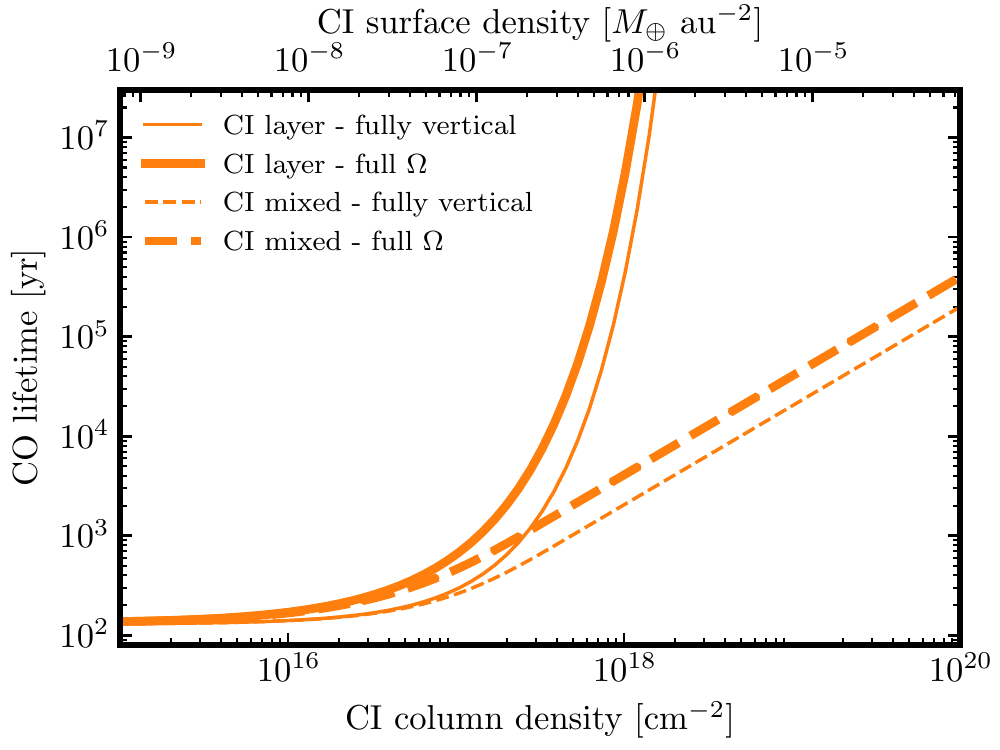}
  \caption{CO lifetime as a function of the CI vertical column density
    in the absence of CO shielding. The solid lines show the average
    lifetime when CI is in two surface layers above and below the CO
    gas. The dashed lines show a case where CO and CI are well
    mixed. The thin and thick lines correspond to considering ISRF
    photons enter the disc only vertically or from all directions,
    respectively.}
  \label{fig:tCO_ci}
\end{figure}

Figure~\ref{fig:tCO_ci} shows the CO average lifetime as a function of
the CI column density in the absence of CO shielding. The solid lines
correspond to a case where the CI gas is on a thin layer above and
below all the CO gas \citep[as assumed in][]{Kral2019, Marino2020gas},
whereas the dashed lines show a case where CI and CO are well mixed
\citep[as assumed in][]{Cataldi2020}. We find that the assumption on
how CI is distributed has profound implications for the lifetime of CO
gas, as pointed out by \citet{Cataldi2020}. If CO and CI are well
mixed, the average lifetime of CO increases only linearly with the CI
column density as the photodissociation layer moves higher up in the
disc\footnote{This can be demonstrated using Equation 9 in
\cite{Cataldi2020} that shows that the photodissociation rate is
proportional to $1-\exp(-\tau)$ and $\tau_{\rm CO}/\tau$. For
$\tau_{\rm CO}\ll\tau_{\rm CI}$ and $\tau_{\rm CI}\gg1$, the CO
photodissociation rate becomes inversely proportional to $\tau_{\rm
  CI}$.}. In fact, we find that self-shielding is more effective than
CI-shielding for the same column density by a factor ${\sim}1.5$. This
strong dependency of the CO lifetime on the CI vertical distribution
means that it is crucial to study the vertical structure of the gas
both theoretically and observationally to advance in our understanding
of the gas evolution.

The model presented above does not include the photodissociating UV
photons from the star. These can dominate the UV radiation field at
100~au for early A-type stars. Nevertheless, \citet{Marino2020gas}
showed that for the typical CO-rich debris discs where gas is
detected, this component is negligible at tens of au due to the high
column densities in the radial direction. Therefore, we omit it from
our calculations.

\subsection{C ionisation and recombination}

The ionisation rate per CI atom is calculated as
\begin{equation}
  R_{\rm ion}(z) = \frac{R_{\rm ion, 0}}{2} \int_{0}^{\upi} \sin(\theta) e^{-N^{\pm}_{\rm CI}\sigma_{\rm CI}/|\cos(\theta)|} \diff \theta,  \label{eq:Rion}
\end{equation}
where $R_{\rm ion, 0}$ is ionisation rate per CI atom in the optically
thin regime
\begin{equation}
  R_{\rm ion, 0} = \int \phi_{\lambda} \sigma_{ \rm CI}\ \diff \lambda \label{eq:Rion0}, \\
\end{equation}
and corresponds to an ionisation timescale of 100~yr. Recombination is
calculated as
\begin{equation}
R_{\rm rc}(z) = \alpha_{\rm rc}(T)n_{\rm e^{-}}(z), \label{eq:Rrc}
\end{equation}
where $\alpha_{\rm rc}(T)$ is the recombination rate of CII that is
dependent on the temperature and is taken from
\citet{Badnell2006}. The number density of electrons $n_{\rm
  e^{-}}(z)$ is set equal to the number density of CII. In other
words, we assume that electrons only originate from CI ionisation and
that CII and electrons are co-located. Finally, the net ionisation
rate is given by
\begin{equation}
  \dot{\rho}_{\rm ion}(z) = \rho_{\rm CI}(z) R_{\rm ion}(z) - \rho_{\rm CII}(z)R_{\rm rc}(z). \label{eq:rhoion}
\end{equation}

\subsection{Viscous evolution}
\label{sec:visevol}
As the gas viscously evolves, angular momentum is transported outwards
while gas mass is mostly moved inwards. This process effectively
removes gas from the centre of the belt forming an accretion disc
interior to the belt and a decretion disc exterior to it. This process
has been studied in detail using 1D simulations that model the radial
evolution of gas \citep[e.g.][]{Kral2016, Moor2019,
  Marino2020gas}. Here, we use a similar approach to \citet{Kral2019}
and \citet{Cataldi2020}, and we approximate the viscous removal of gas
at the belt central radius with
\begin{equation}
  \dot{\rho}_{\rm vis, i}(z) = \frac{\rho_i(z)}{t_{\rm vis}}, \label{eq:rhovis}
\end{equation}
where $t_{\rm vis}$ is a viscous timescale. An expression for $t_{\rm
  vis}$ can be found by considering that in steady-state, the gas
accretion rate \citep[$3\upi \nu \Sigma$,][]{Armitage2011} must be
equal to the gas input rate $\dot{M}_{+}$, and that $\dot{\Sigma}_{+}$
must be equal to $\Sigma/t_{\rm vis}$. Combining these two equalities,
we find
\begin{equation}
  t_{\rm vis} = \frac{\sqrt{\upi} r \Delta r }{3 \sqrt{2\ln(2)} \nu} \approx \frac{r \Delta r }{2 \nu}.  \label{eq:tvis}
\end{equation}
With this definition, the steady-state surface density
($\dot{M}_{+}/(3\upi\nu)$) is independent of $\Delta r$, and matches
the surface density obtained solving the radial viscous evolution and
the analytic solution presented in Equation B13 in Metzger+2012 (valid
when $\Delta r \xrightarrow{}0$). Note that when approximating the
viscous evolution through Equation~\ref{eq:rhovis} (only truly valid
at steady-state), the densities tend to grow faster and steady state
is reached earlier than if we were to solve the viscous evolution
radially. Finally, the kinematic viscosity is parametrized with the
standard $\alpha$ viscosity as $\nu = \alpha c_{\rm s} H$
\citep{Shakura1973}.

\subsection{Gas vertical diffusion}
\label{sec:diffusion}

The last effect to consider is the vertical diffusion, which will mix
the different species. The diffusion term in Equations \ref{eq:CO},
\ref{eq:CI}, and \ref{eq:CII} can be written as \citep{Ilgner2006,
  Armitage2011}
\begin{align}
  & \dot{\rho}_{i, D} = \frac{\partial}{\partial z} \left[\rho D \frac{\partial}{\partial z} \left(\frac{\rho_i}{\rho}\right) \right], \label{eq:diff} \\
\end{align} 
where $D$ is the mass diffusivity, which we assume arises from
turbulent mixing. Hence, we set $D$ equal to the vertical kinematic
viscosity, $\nu_{\rm v}$, which is parametrised as $\alpha_{\rm v}
c_{\rm s} H$. We will assume $\alpha_{\rm v}=\alpha$ unless otherwise
stated, i.e. the vertical diffusivity is as strong as the radial
kinematic viscosity. This means that the viscosity is isotropic, but
it is possible that the vertical and radial viscosity and diffusion
could differ (see discussion in \S\ref{dis:diffusion}). Therefore, we
also explore scenarios where $\alpha_{\rm v}$ is independent of
$\alpha$. Finally, we can define the vertical diffusion timescale as
$t_{\rm diff}=H^2/D=1/(\alpha_{\rm v} \Omega_{\rm K})$, which
translates to
\begin{equation}
  t_{\rm diff}= 1.3\times10^{4}\ \mathrm{yr}\ \left(\frac{\alpha_{\rm v}}{10^{-2}}\right)^{-1} \left(\frac{r}{100\ \mathrm{au}}\right)^{3/2} \left(\frac{M_\star}{1.5\ M_{\odot}}\right)^{-1/2}. \label{eq:tdiff}
\end{equation}




\section{Results}
\label{sec:results}
With the model presented above, we now proceed to simulate its
evolution. We consider a system around a 1.5~$M_{\odot}$ star with a
luminosity of 10~$L_{\odot}$. The system has a planetesimal belt
centred at 100~au, with a FWHM of 50~au. We assume an alpha viscosity
of $10^{-2}$, which translates to a viscous timescale of 1.5 Myr as
defined in Equation~\ref{eq:tvis}. Note that although the viscosity
parameter is kept constant, below we present the evolution for
different values of $\alpha_{\rm v}$ and different gas input rates to
cover the different scenarios. Our vertical grid consists of only one
hemisphere since the disc is symmetric with respect to the midplane,
and is divided into 50 bins from the midplane to 5 scale heights. The
polar angle $\theta$ is sampled with 20 bins from 0 to $\pi$. We
evolve the system for 10~Myr.

\subsection{Without diffusion}
\label{sec:without_diffusion}
\begin{figure*}
  \centering \includegraphics[trim=0.0cm 0.0cm 0.0cm 0.0cm, clip=true,
    width=0.9\textwidth]{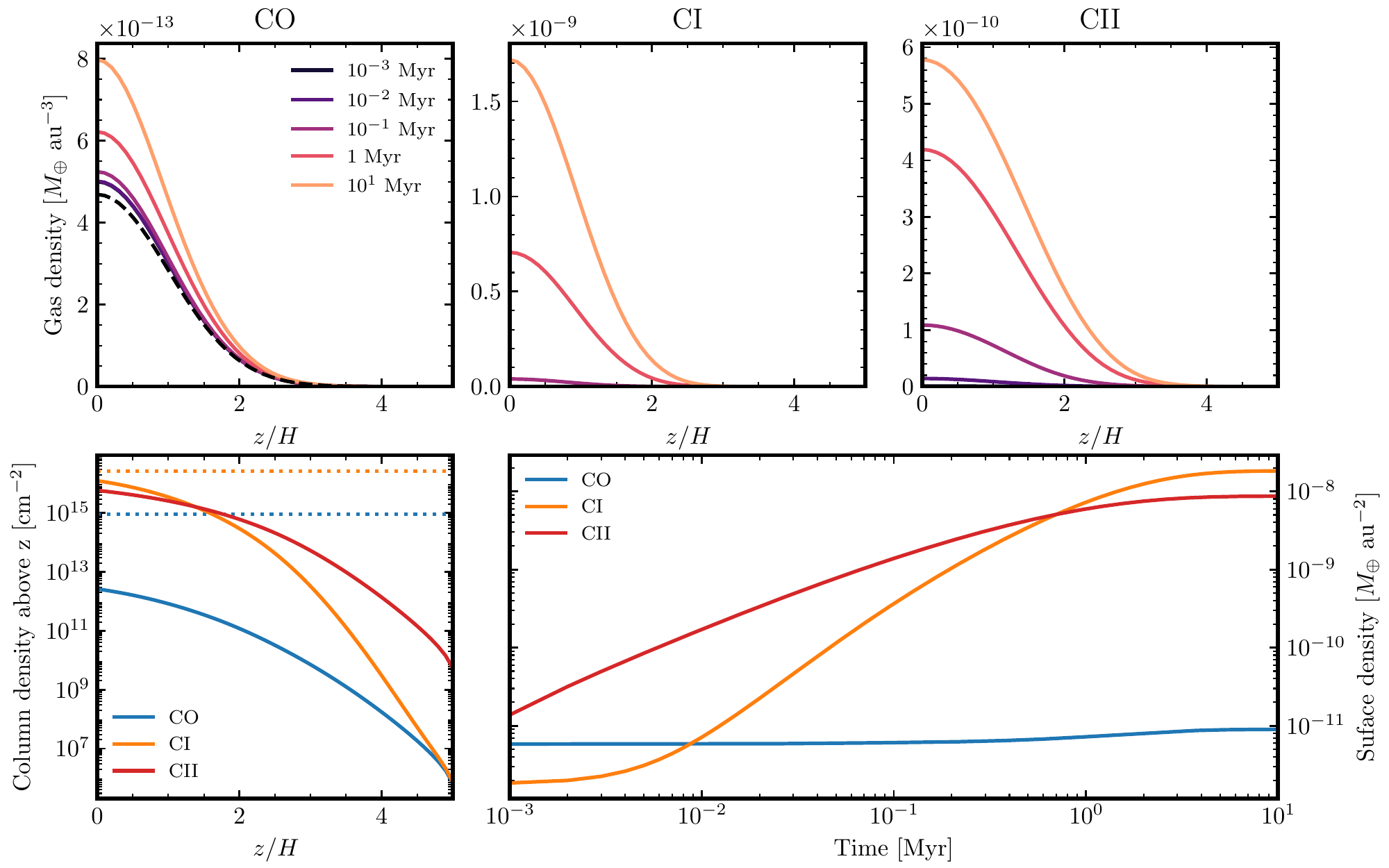}
  \caption{Evolution of CO, CI and CII for
    $\dot{M}_{+}=10^{-3}$~$M_{\oplus}$~Myr$^{-1}$, $\alpha=10^{-2}$
    and no vertical diffusion. The top panels show the vertical
    density profile at different epochs as a function of $z/H$. The
    black dashed line in the top left panel shows the expected CO
    density if there was no shielding. The bottom left panel shows the
    vertical column density above $z$ as a function of $z/H$ (the
    total column density is twice the value reached at $z=0$) for CO
    (blue), CI (orange) and CII (red). The horizontal dotted lines
    show the critical column density of CO (blue) and CI (orange) to
    shield the CO underneath it by a factor $\geq e$. The bottom right
    panel shows the evolution of the surface density of CO, CI and
    CII.}
  \label{fig:lowmdot_nodiff}
\end{figure*}

Figure \ref{fig:lowmdot_nodiff} shows a case in which CO gas is input
at a rate of $10^{-3}$~$M_{\oplus}$~Myr$^{-1}$ and diffusion is
switched off. The top panels show the vertical distribution of CO, CI,
and CII at five different epochs. The three species have a vertical
distribution that approximates a Gaussian. The bottom left panel shows
the column density as a function of height. The horizontal dotted
lines show the critical column density to shield the CO (below $z$)
via self-shielding (blue) and CI-shielding (orange) by a factor $\geq
e$. From this we can see that CO is only marginally shielded by CI as
the column density of CI at the midplane is close to the critical
column density. The bottom right panel shows the evolution of the
surface density of the three species. This shows how the surface
density of CO stays roughly constant at a value where the
photodissociation rate is balanced by the CO input rate. While CO
photodissociates, carbon is produced and quickly ionised in a 100~yr
timescale. Ionised carbon dominates the first Myr of evolution, while
CI dominates afterwards as recombination becomes more effective within
1.5 scale heights. After 3~Myr, the CI gas is close to steady state at
surface density of $2\times{10^{-8}}\ M_{\oplus}$~au$^{-2}$. At this
level CO is marginally shielded and its surface density increases by
50\%. This case represents the systems with low CO masses where CO is
released by planetesimals and destroyed in timescales of
${\sim}100$~yr like in $\beta$~Pic, HD~181327, and Fomalhaut
\citep{Dent2014, Marino2016, Matra2017fomalhaut}.

\begin{figure*}
  \centering \includegraphics[trim=0.0cm 0.0cm 0.0cm 0.0cm, clip=true,
    width=0.9\textwidth]{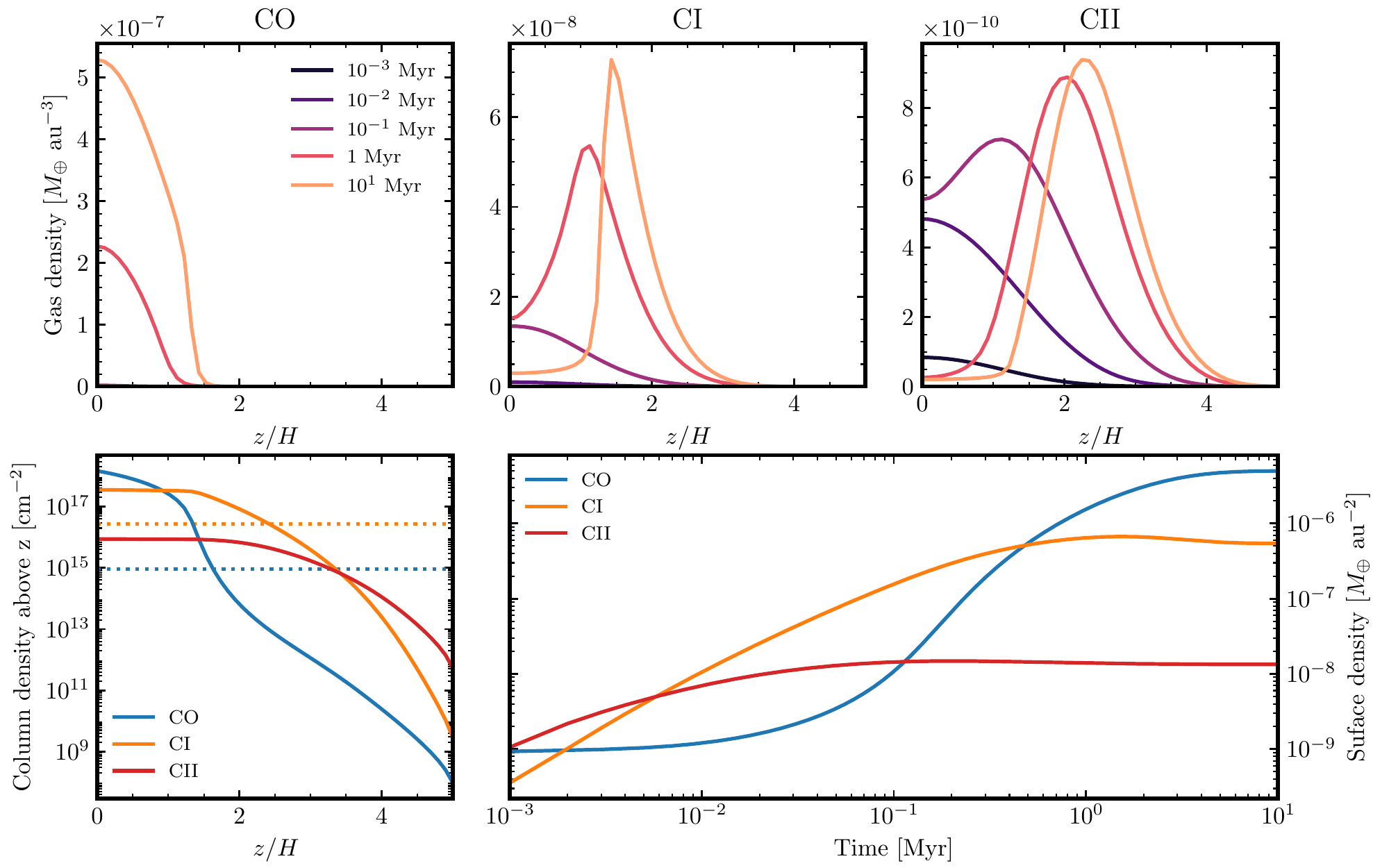}
  \caption{Same as Figure~\ref{fig:lowmdot_nodiff}, but for
    $\dot{M}_{+}=0.1$~$M_{\oplus}$~Myr$^{-1}$, $\alpha=10^{-2}$
    and no vertical diffusion. }
  \label{fig:highmdot_nodiff}
\end{figure*}

Figure \ref{fig:highmdot_nodiff} shows a case with a hundred times
higher CO release rate ($\dot{M}_{+}=0.1$~$M_{\oplus}$~Myr$^{-1}$). In
this case CO becomes shielded after 1~Myr by CI and it reaches a
photodissociation timescale of 6~Myr by the end of the
simulation. This timescale is longer than the viscous timescale and
thus CO is mainly lost via viscous evolution. As CI becomes optically
thick to UV radiation, it forms a layer at $1.5 H$ where CI production
peaks, i.e. $\dot{\rho}_{\rm ph}(z)-\dot{\rho}_{\rm ion}(z)$ reaches
its maximum. CII is mostly produced above this layer at $2H$ where
carbon is still produced via photodissociation of new CO, and
ionisation and recombination rates are comparable. This evolution
confirms that it is possible that CI is in a layer above the midplane
that is dominated by CO, as assumed in \citet{Kral2019} and
\cite{Marino2020gas}. This shielded case would correspond to the
gas-rich discs, where CO masses and surface densities are found to be
comparable to protoplanetary disc levels. Note that the height at
which these layers are (relative to the scale height) will depend on
the rate at which CO gas is released. The higher the release rate is,
the higher these layers will be (see \S\ref{sec:vert}).

\subsection{With diffusion}
\label{sec:results_diff}

Now we focus on the effect of vertical diffusion. Vertical diffusion
happens on a timescale of $10^{4}$~yr for $\alpha=10^{-2}$
(Equation~\ref{eq:tdiff}). It is worth noting that since $H$ is
typically an order of magnitude smaller than $r$, vertical diffusion
can be two orders of magnitude faster than viscous evolution. Vertical
diffusion is, nevertheless, longer than CO photodissociation and
ionisation timescales in the unshielded case.  We thus do not see
differences between a case with diffusion and without when the CO
release rate is low, as in the case shown in
Figure~\ref{fig:lowmdot_nodiff}. On the other hand, if the gas release
rate is high both ionisation and photodissociation rates become longer
than the diffusion timescale, and thus CO and CI can easily mix.

\begin{figure*}
  \centering \includegraphics[trim=0.0cm 0.0cm 0.0cm 0.0cm, clip=true,
    width=0.9\textwidth]{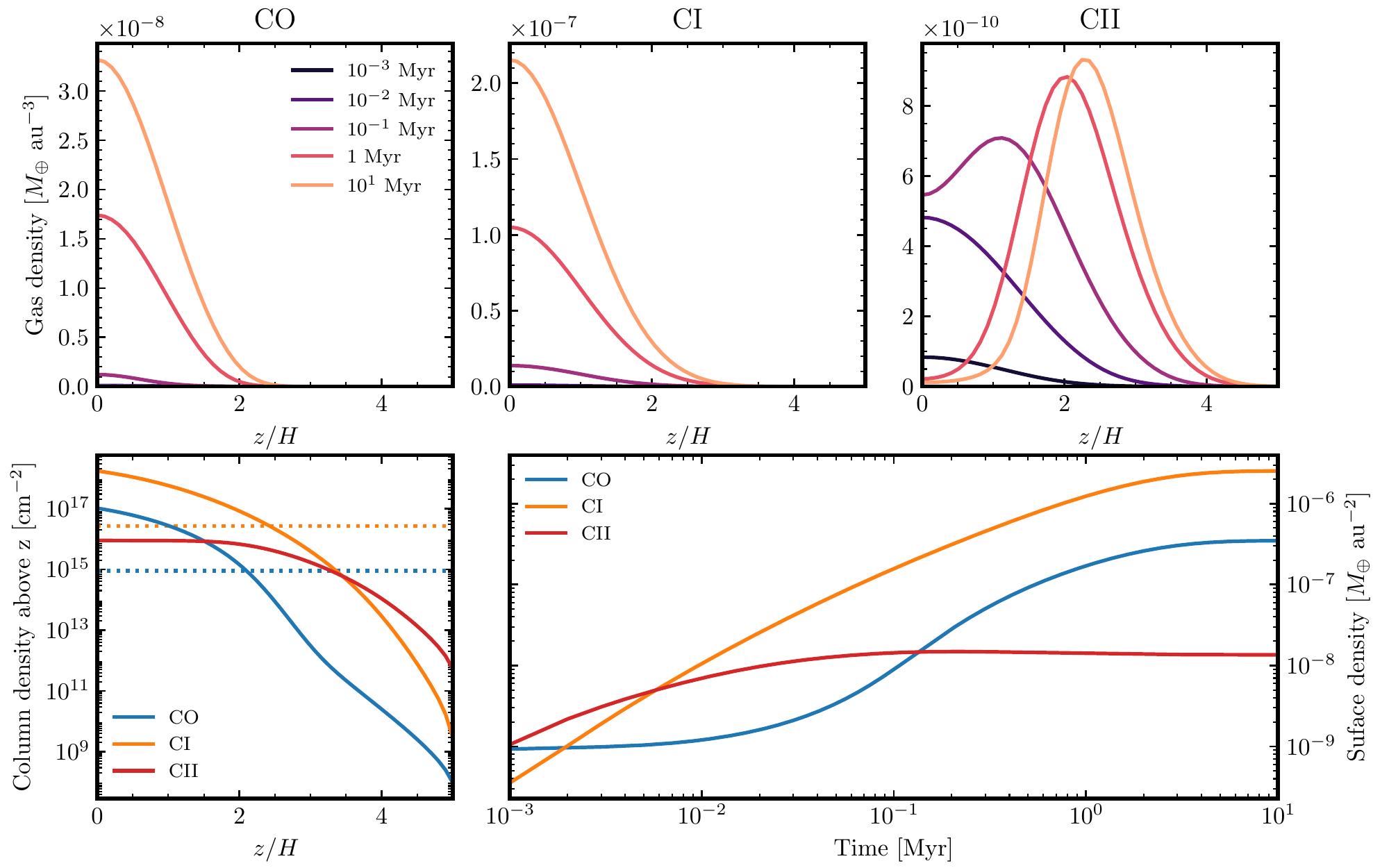}
  \caption{Same as Figure~\ref{fig:lowmdot_nodiff}, but for
    $\dot{M}_{+}=0.1$~$M_{\oplus}$~Myr$^{-1}$, $\alpha=10^{-2}$
    and with vertical diffusion.}
  \label{fig:highmdot_diff}
\end{figure*}

Figure~\ref{fig:highmdot_diff} shows this latter scenario where CO gas
is released at a high rate of $0.1$~$M_{\oplus}$~Myr$^{-1}$. Diffusion
mixes CO and CI, moving CO to upper layers where it becomes exposed
and CI is transported towards the midplane reducing its effectiveness
at shielding CO. Both CO and CI show a vertical distribution close to
Gaussian with a standard deviation similar to $H$. Given the less
effective shielding by CI, the photodissociation timescale drops to
0.09 Myr (almost a factor 100 shorter than the case without
diffusion). The CO surface density, however, drops by a smaller factor
close to 10. This difference is because in the scenario without
diffusion, CO was being lost via viscous evolution (i.e. in a viscous
timescale of ${\sim}1$~Myr), while with diffusion it is being lost via
photodissociation. The CII distribution remains unaffected since carbon
ionisation and recombination occur on a shorter timescales than
diffusion.

\begin{figure*}
  \centering \includegraphics[trim=0.0cm 0.0cm 0.0cm 0.0cm, clip=true,
    width=0.9\textwidth]{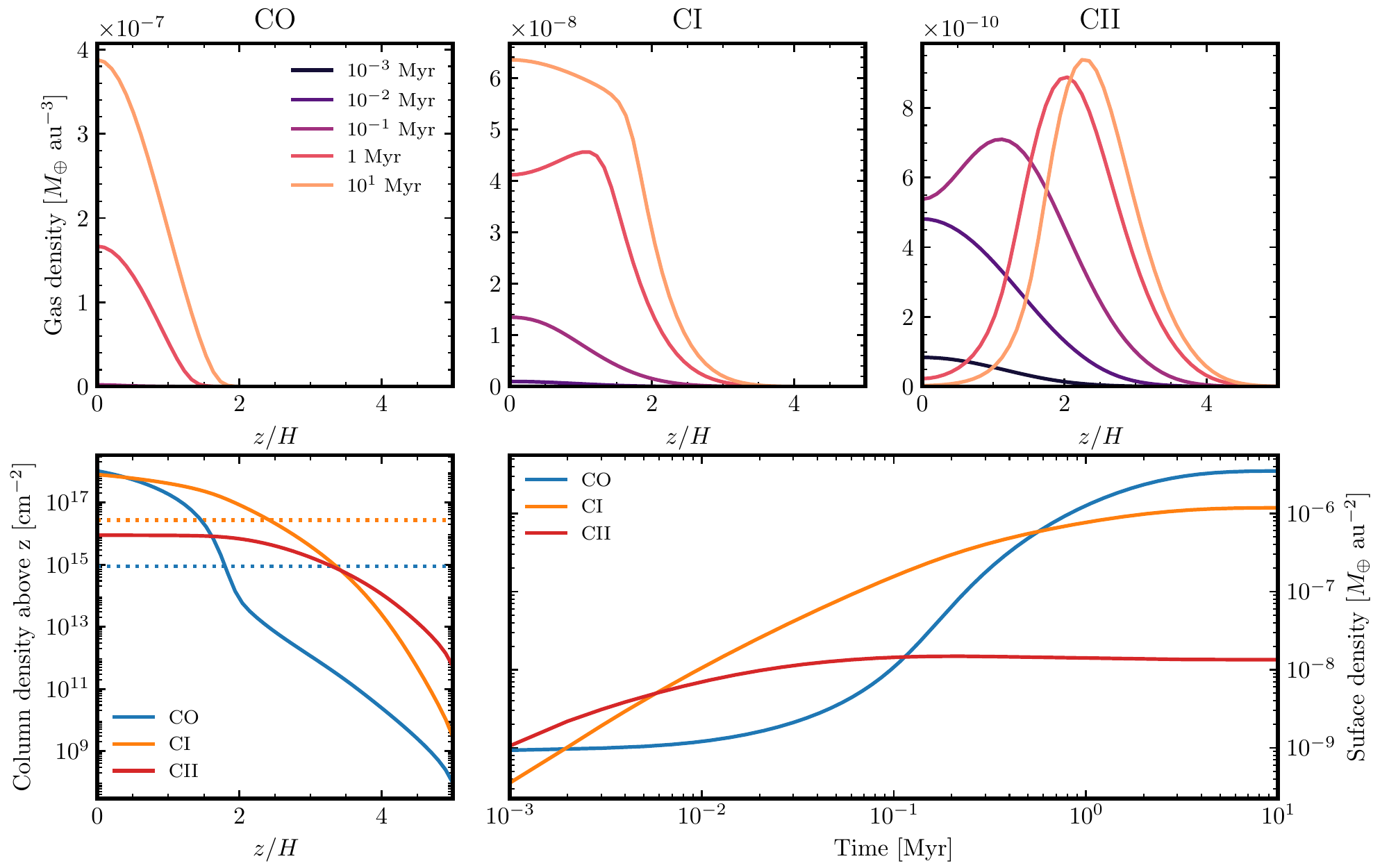}
  \caption{Same as Figure~\ref{fig:lowmdot_nodiff}, but for
    $\dot{M}_{+}=0.1$~$M_{\oplus}$~Myr$^{-1}$, $\alpha=10^{-2}$ and
    with a vertical diffusion strength of $\alpha_{\rm v}=10^{-4}$.}
  \label{fig:highmdot_lowdiff}
\end{figure*}

It is also possible to consider a case where the vertical and radial
values of $\alpha$ differ as mentioned in \S\ref{sec:diffusion}. In
Figure~\ref{fig:highmdot_lowdiff} we show a case where vertical
diffusion (parametrized now by $\alpha_{\rm v}$) is weaker and equal
to $10^{-4}$. Since $H$ is a factor ${\sim}10$ smaller than the radius
and width, the now hundred times weaker vertical diffusion leads to a
vertical diffusion timescale similar to the viscous timescales and
close to 1~Myr. At this critical value, mixing is significant causing
the CI density to peak at the midplane but slow enough for CI to have
a wide vertical distribution. Compared to the case with $\alpha_{\rm
  v}=\alpha=10^{-2}$, CO is better shielded and reaches a surface
density similar to the case with no vertical diffusion. Therefore, we
conclude that for $\alpha_{\rm v}< \alpha (H/r)^2$ vertical diffusion
plays a minor role. Conversely, if $\alpha_{\rm v}> \alpha (H/r)^2$
and the diffusion timescale is shorter than the age of the system,
vertical diffusion mixes CO and CI. For systems younger than the
diffusion timescale ($\sim10^2/\alpha_{v}$ yr at 100~au,
Equation~\ref{eq:tdiff}), or systems in which gas has been released
for a shorter timescale, vertical mixing will not be able to
effectively mix CO and CI.

\subsection{The vertical distribution of CO and CI}
\label{sec:vert}

\begin{figure}
  \centering \includegraphics[trim=0.0cm 0.0cm 0.0cm 0.0cm, clip=true,
    width=1.0\columnwidth]{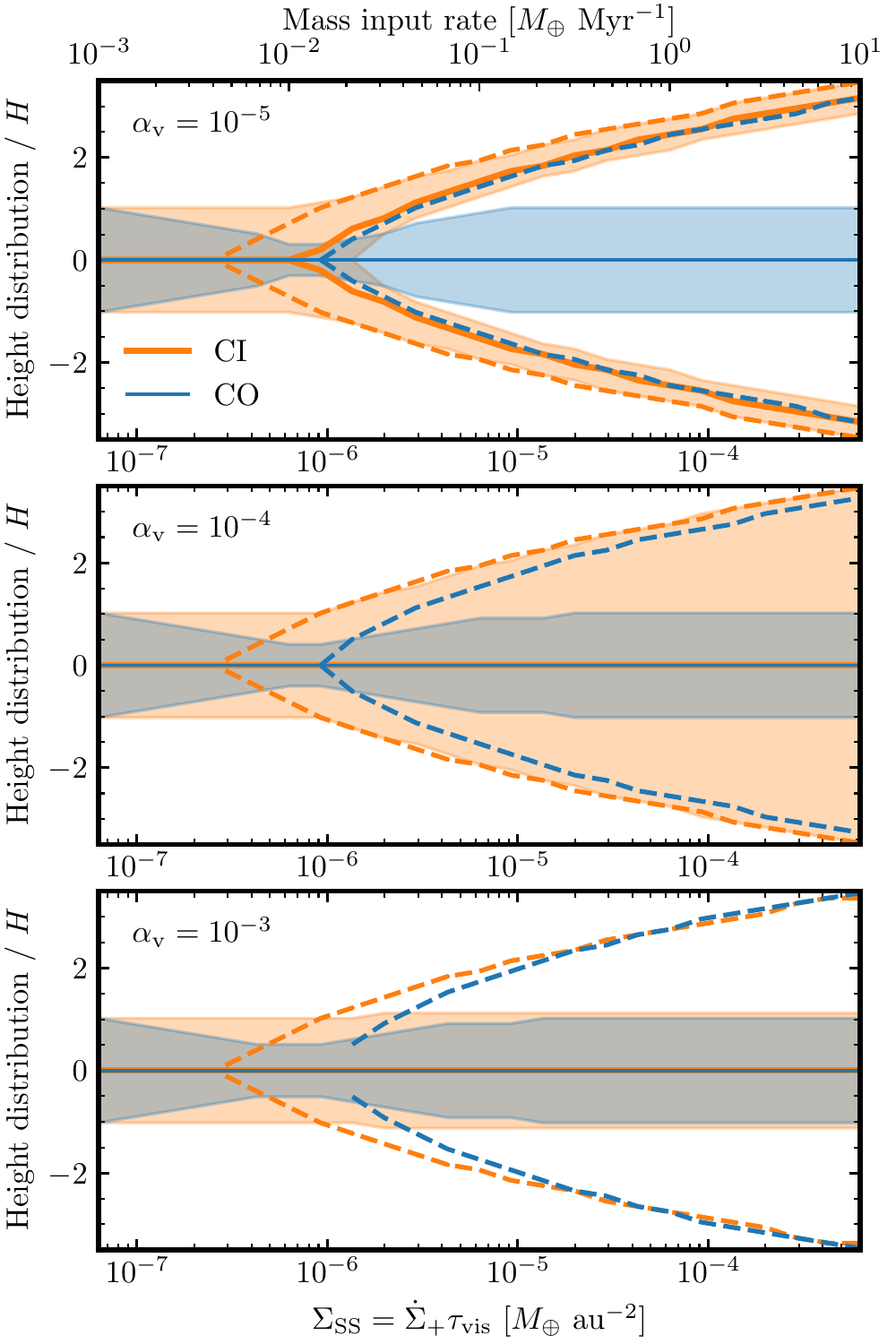}
  \caption{Vertical distribution in steady state of CO (blue) and CI
    (orange) normalized by the scale height ($H$) as a function of the
    steady state surface density of gas, in a case with
    $\alpha=10^{-2}$. The solid line shows where the density
    distribution peaks, while the shaded region represents the region
    where the density is at least 60\% of the peak density ($1\sigma$
    drop). The dashed lines represent the height at which CI and CO
    become optically thick to photodissociating radiation in the
    vertical direction. The top, middle, and bottom panels show
    scenarios where $\alpha_{\rm v}$ is $10^{-5}$, $10^{-4}$, and
    $10^{-3}$, respectively. The upper x-axis in the top panel shows
    the mass input rates as reference. Note that the orange solid line
    in the middle and bottom panels lies behind the solid blue line. }
  \label{fig:vertical_distribution}
\end{figure}

In order to understand how the vertical distribution of CO and CI
depend on the gas input rate and $\alpha_{\rm v}$ we run
\textsc{exogas} for a sample of different values fixing
$\alpha=10^{-2}$ and evolving the system until it reaches steady
state. The main effect of changing the gas input rate is to change the
steady state surface density of gas ($\Sigma_{\rm SS}$), which is
simply the product of the gas input rate ($\dot{\Sigma}_{+}$ as
defined in Equation~\ref{eq:sigmap}) and the viscous
timescale. Therefore, below we will focus on how the CO and CI
distribution change as a function of this steady state surface
density. Figure~\ref{fig:vertical_distribution} shows where the CO
(blue) and CI (orange) densities peak (solid lines) as a function of
the steady state surface density. The shaded area represents the
region where the density is at least 60\% of the peak density
(equivalent to a $\pm H$ range for a Gaussian distribution), and the
dashed lines the height at which CO and CI become optically thick to
photodissociating radiation in the vertical direction (i.e. the
$\tau=1$ layer for each species).

The top panel shows cases with low vertical diffusion ($\alpha_{\rm
  v}=\alpha/1000=10^{-5}$). For low gas densities, CO and CI have the
same vertical distribution as shown in
Figure~\ref{fig:lowmdot_nodiff}. As the surface density increases, we
find an intermediate stage where CI gas becomes optically thick and
can shield CO, which becomes highly peaked at the disc midplane. At
higher surface densities, the CI density peaks in a surface layer at a
height just below its $\tau=1$ layer, which increases with the surface
density. This layer traces the location where the CO photodissociation
rate is the highest. As CO is well shielded under this layer, the CO
vertical extent becomes wider, closer to a Gaussian and with a
standard deviation equal to $H$.

The middle panel shows cases with an intermediate vertical diffusion
($\alpha_{\rm v}=\alpha/100=10^{-4}$). The behaviour is similar to the
one with low vertical diffusion, e.g. the CO vertical distribution and
$\tau=1$ layer are almost identical. However, the CI distribution now
peaks at the midplane, still extending to heights similar to the case
with low vertical diffusion. This intermediate case demonstrates how
even if the CI density peaks at the midplane, the CI equivalent scale
height can be much larger than the true gas scale height.

The bottom panel shows a case with a high vertical diffusion
($\alpha_{\rm v}=\alpha/10=10^{-3}$). In this case, both CO and CI
have an almost identical vertical distribution, except in the
intermediate stage when CI starts to become optically thick. In this
range of surface densities, CO is moderately shielded near the
midplane and thus its density strongly peaks at $z=0$. Note that the
diffusion timescale is not short enough compared to the
photodissociation timescale to smear out this peak. Only for
$\alpha_{\rm v}{\sim}1$, the vertical diffusion timescale would become
${\sim}100$~yr, i.e.  comparable to the CO photodissociation timescale,
and thus short enough to widen the vertical distribution of CO in this
range of surface densities. Note that a stronger vertical diffusion
(e.g. $\alpha_{\rm v}=\alpha=10^{-2}$) leads to very similar results
where CO and CI are well mixed. The main difference is that at the
intermediate stage the vertical spread of CO is slightly wider.

Finally, although the example shown in
Figure~\ref{fig:vertical_distribution} shows a case with
$\alpha=10^{-2}$, we find similar results for higher and lower
$\alpha$'s if we keep $\alpha_{\rm v}/\alpha$ constant. An example of
this is shown in the Appendix~\ref{a:vertical}, where we show that
with $\alpha=10^{-1}$ the same vertical structures are obtained. In
other words, for a fixed $H/r$ whether CI will form a surface layer or
will be bound to the midplane will ultimately depend on the ratio
$\alpha_{\rm v}/\alpha$, the viscosity (or gas removal timescale) and
the gas production rate. As mentioned in \S\ref{sec:results_diff}, in
systems younger than the vertical diffusion timescale
(Equation~\ref{eq:tdiff}), i.e. not yet in steady state, the CI could
form a surface layer regardless of the ratio $\alpha_{\rm v}/\alpha$
if the gas surface density is high enough for CI to become optically
thick in the vertical direction.

\subsection{The surface density of CO and CI}

\begin{figure}
  \centering \includegraphics[trim=0.0cm 0.0cm 0.0cm 0.0cm, clip=true,
    width=1.0\columnwidth]{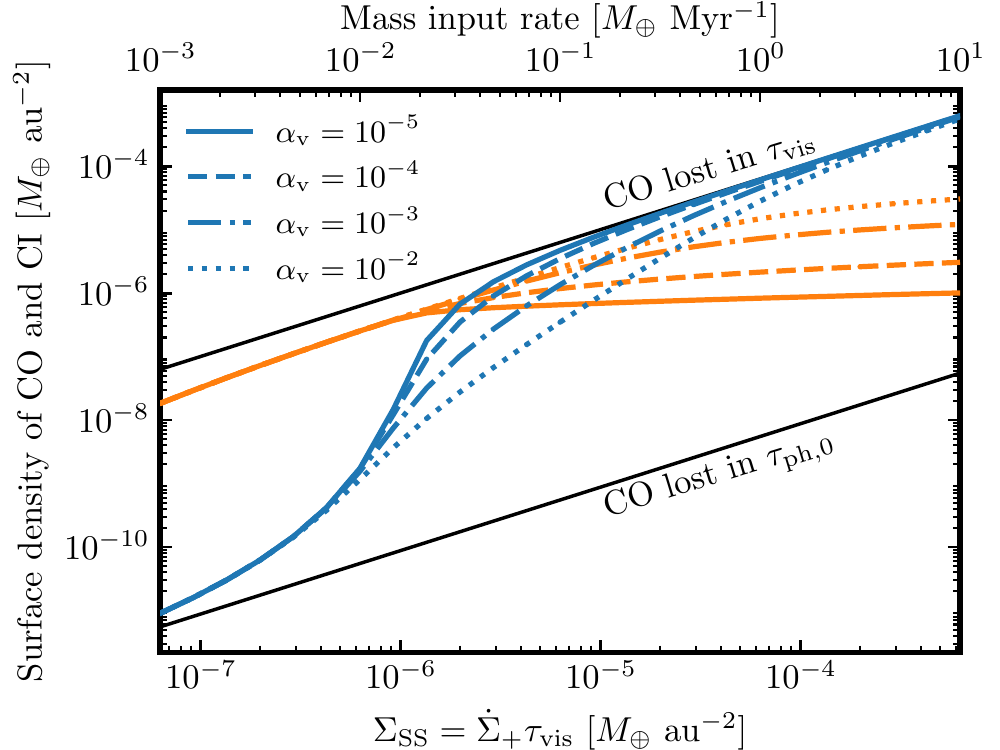}
  \caption{Surface density of CO (blue) and CI (orange) as a function
    of the steady state surface density of gas, in a case with
    $\alpha=10^{-2}$. The solid, dashed, dot-dashed and dotted lines
    represent cases with $\alpha_{\rm v}=10^{-5}$, $10^{-4}$,
    $10^{-3}$, and $10^{-2}$, respectively. The upper and lower black
    solid lines represent the maximum and minimum surface density of
    CO, i.e. when it is lost in a viscous timescale or unshielded and
    lost in a 130~yr photodissociation timescale. The upper x-axis
    shows the mass input rates as reference.}
  \label{fig:density_vs_alphavs}
\end{figure}

With a better understanding of how the vertical distribution of CI
depends on $\alpha_{\rm v}$, we now focus on how the surface density
of CO and CI will depend on $\alpha_{\rm
  v}$. Figure~\ref{fig:density_vs_alphavs} shows the surface density
of CO (blue) and CI (orange) as a function of the steady state surface
density of gas, for different values of $\alpha_{\rm v}$, assuming
$\alpha=10^{-2}$. As expected, the lower $\alpha_{\rm v}$ is, the
better CI is at shielding CO as it is mostly in a surface layer that
absorbs most UV photons. The upper and lower black diagonal lines show
the expected surface density of CO if it is lost through viscous
evolution or if it is unshielded and lost through photodissociation,
respectively. For a low $\alpha_{\rm v}$ (solid line), as $\Sigma_{\rm
  SS}$ increases the CI production rate stalls as most UV photons are
absorbed by CI rather than CO. This is why the surface density of CI
is almost constant for $\Sigma_{\rm
  SS}\geq10^{-6}\ M_{\oplus}$~au$^{-2}$. This also causes the CO
lifetime to increase exponentially up to a viscous timescale, and the
surface density quickly converges to the upper black line.

On the other hand, as described before, with a high value of
$\alpha_{\rm v}$ (dotted line) CI is well mixed with CO and thus it is
not as effective at shielding it from photodissociating
radiation. Thus, for the same $\Sigma_{\rm SS}$ or gas input rate, the
CO surface density is much lower than in the case with a weak vertical
diffusion. This is effectively the result obtained by
\cite{Cataldi2020}.

The difference between the blue curves is the largest near
$\Sigma_{\rm SS}=2\times10^{-6}\ M_{\oplus}$~au$^{-2}$, where the
surface density of CO for $\alpha_{\rm v}=10^{-5}$ is ${\sim}30$ times
larger than for $\alpha_{\rm v}=10^{-2}$. How big this difference is
also depends on $\alpha$. The lower the viscosity, the larger this
difference becomes as shielding can extend the CO lifetime to an even
longer viscous timescale.

\section{Discussion}
\label{sec:discussion}

In this section we discuss the implications of our findings, what is
known about vertical diffusion in circumstellar discs, and the level
of ionisation of the gas obtained from our simulations. We also
discuss the effect vertical diffusion can have on the vertical
settling of dust and on the radial evolution of CO, how high
resolution observations could constrain the strength of vertical
diffusion, but might be unable to constrain the gas scale height, and
finally, we discuss some of the limitations of our model.

\subsection{Implications for shielded/hybrid discs}

As pointed out by \cite{Cataldi2020} and demonstrated in detail here,
if CO and CI are well mixed shielding is much less effective than if
CI is in a surface layer. To achieve the same level of shielding, much
higher quantities of CI gas are needed. However, CI observations of
gas-rich discs seem to suggest lower CI quantities than those needed
if CO and CI are well mixed. Two scenarios might solve this
tension. On one hand, \cite{Cataldi2020} suggested that if CI was
captured by dust grains and then used to re-form CO that is then
released, this could maintain a high CO production rate in the
system. In this recycling scenario, CO would be predominately
self-shielded. This scenario, however, still needs to be tested via
physical simulations or experiments. On the other hand, the gas could
be primordial or hybrid in origin, containing still large amounts of
hydrogen that shield the CO \citep{Kospal2013}. Note that although the
CI mass has been constrained, its value is still largely dependent on
assumptions such as its temperature and vertical structure that can
change its optical depth when viewed close to edge-on. Therefore, it
is possible that the real mass of CI in the gas-rich discs is much
larger than the current observational estimates.

If diffusion is weak, nevertheless, we expect the CI density to peak
above and below the CO-dominated midplane, as implicitly assumed in
the simple shielded secondary scenario presented by \cite{Kral2019}. A
weak diffusion could thus explain the population of gas-rich discs and
account for the current CI mass estimates. Unfortunately, there are no
clear predictions for how strong vertical diffusion will be (see
\S\ref{dis:diffusion} below). This will ultimately depend on the
relative strength of the turbulence in the vertical and radial
direction. Therefore, high-resolution ALMA observations are crucial
since they reveal the true nature of gas-rich debris discs and provide
constraints on the hydrodynamics (see \S\ref{sec:observability}).

\subsection{Vertical diffusion}
\label{dis:diffusion}
Since the strength of the vertical diffusion in debris discs is not
well known, we may look for some guidance from the study of
protoplanetary discs, keeping in mind that the gas densities and
especially the dust optical depths are significantly higher than in
debris discs. One way to constrain the vertical diffusion in
protoplanetary discs is by considering the disc chemistry. For
example, in order to explain the high CH$_3$CN abundance observed
towards MWC~480, the disc model by \citet{Oberg2015} requires
efficient vertical transport ($\alpha_{\rm v}{\sim}10^{-3}$--$10^{-2}$)
of CH$_3$CN ice from the midplane to UV-exposed regions in the disc
atmosphere where the ice can be photodesorbed. Similarly efficient
vertical mixing is also believed to contribute to the observed C
depletion in protoplanetary discs by transporting CO from the disc
atmosphere to the midplane where it can freeze out and be sequestered
into pebbles \citep[e.g.][]{Kama2016,VanClepper2022}. Dust settling
has also been used to constrain vertical diffusion. For example,
\citet{Pinte2016} report that their dust settling model best fits the
small dust scale height of the HL~Tau protoplanetary disc if weak
vertical diffusion is assumed ($\alpha_{\rm v}$ a few times
$10^{-4}$).

Recent observations aimed at detecting non-thermal gas motion form CO
line emission suggest that relatively weak turbulence (corresponding
to $\alpha\lesssim10^{-3}$) may be common in protoplanetary discs
\citep{Flaherty2018,Flaherty2020}. In general, it is often assumed
that $\alpha_{\rm v}$ is approximately equal to the turbulent $\alpha$
parameter. This can be justified from numerical simulations of
protoplanetary discs subject to the magneto-rotational instability
(MRI), showing that $\alpha_{\rm v}$ is usually within an order of
magnitude of $\alpha$ \citep[e.g.][]{Okuzumi2011,Zhu2015,Xu2017}. To
the best of our knowledge, the only study looking at the MRI in debris
discs was presented by \citet{Kral2016b}. They showed that the MRI
could be at work in the debris disc around $\beta$~Pictoris, although
a weak magnetic field would be sufficient to stabilise the
disc. However, their results might not be directly applicable to the
CO-rich discs we consider here due to the significantly higher gas
densities and lower ionisation compared to the disc around
$\beta$~Pic (see \S\ref{sec:ion}).

As a further complication, the turbulence strength could vary
vertically as the ionisation level increases with $z$
\citep{Flock2015, Delage2021}. Also, if turbulent motion is driven by
a mechanism different from the MRI, $\alpha$ could be very different
from $\alpha_{\rm v}$. Numerical simulations by \citet{Stoll2017} show that
turbulence induced by the vertical shear instability could be highly
anisotropic ($\alpha_{\rm v}$ more than two orders of magnitude larger than
$\alpha$).

In summary, the strength of vertical diffusion in debris discs is
essentially unknown. Thus, in this study we explored a range of
plausible values for $\alpha_{\rm v}$ to understand the gas evolution
in different regimes. Future observations of highly-inclined gas-rich
debris discs could provide the first constraints on $\alpha_{\rm v}$
and the hydrodynamics (see \S\ref{sec:observability}).

\subsection{Ionisation fraction}
\label{sec:ion}
As discussed above, one possible driver of vertical (and radial)
diffusion in debris discs is turbulence due to the MRI. Well-ionised
debris discs such as $\beta$~Pic may be unstable to the MRI, provided
that the magnetic field is not too strong \citep{Kral2016b}. If,
however, the ionisation fraction is too low, the gas decouples from
the magnetic field and the disc is stable to the MRI \citep[in
  protoplanetary discs, such regions are known as ``dead
  zones'';][]{Gammie1996}.
 
\begin{figure}
  \centering \includegraphics[trim=0.0cm 0.0cm 0.0cm 0.0cm, clip=true,
    width=1.0\columnwidth]{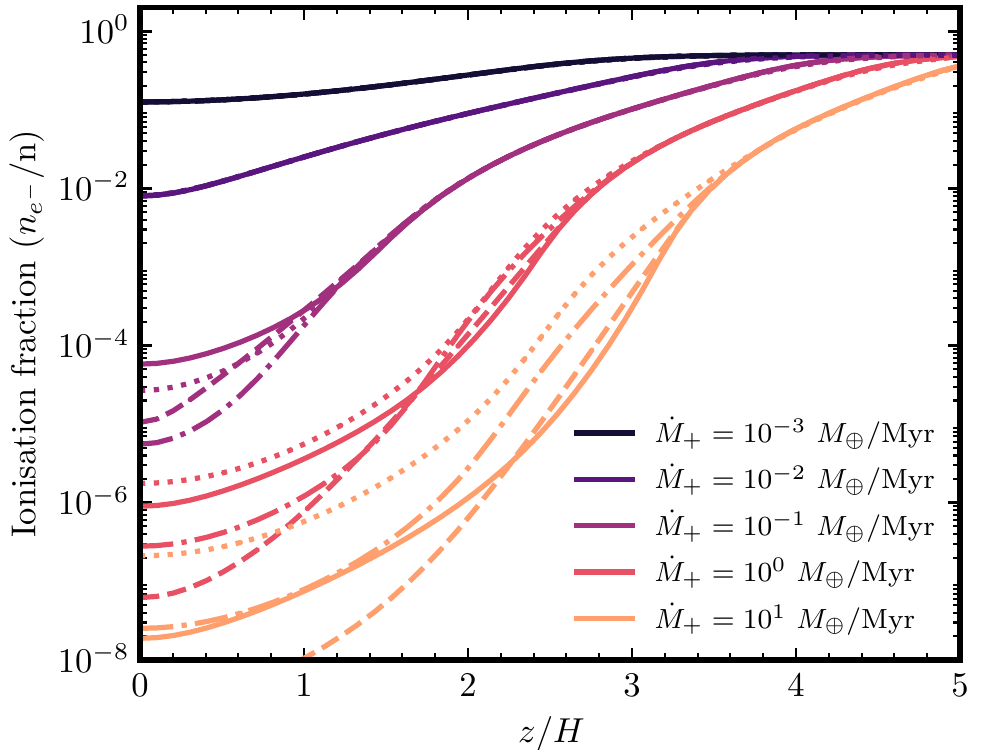}
  \caption{Ionisation fraction as a function of height for a disc with
    $\alpha=10^{-2}$. The different colours correspond to different
    gas input rates. The solid, dashed, dot-dashed, and dotted lines
    represent cases with $\alpha_{\rm v}=10^{-5}$, $10^{-4}$,
    $10^{-3}$, and $10^{-2}$, respectively. The ionisation fraction is
    calculated as the number density ratio between CII (equivalent to
    the electron number density) and CO+CI+CII+O.}
  \label{fig:ion}
\end{figure}

Consideration of the disc vertical structure is important for the
calculation of the ionisation fraction in debris
discs. \citet{Kral2019} found that the ionisation fraction near the
disc midplane can be several orders of magnitude lower than that at
the disc surface, due to the self-shielding of the atomic carbon. This
effect is also present in the models presented in this paper (see
Figure~\ref{fig:ion}), where both the atomic carbon and CO in the
upper disc layers shield the disc midplane from the ISRF. We find
that the ionisation fraction at the disc midplane can vary greatly
with the disc mass (which is, at fixed $\alpha$, proportional to the
gas mass input rate shown in Figure~\ref{fig:ion}). In a more massive
disc, the disc midplane is better shielded and this results in both
lower rate of carbon production and lower ionisation rate of the
produced carbon. Vertical diffusion also influences the midplane
ionisation fraction, to a lesser degree. We find that the relationship
between the degree of vertical diffusion (i.e., $\alpha_{\rm v}$) and
the midplane ionisation fraction is non-monotonous. This is due to the
relative importance of two competing effects. As the vertical
diffusion coefficient first increases from $10^{-5}$ to $10^{-4}$
(solid to dashed line), diffusion is strong enough for CI from upper
layers to diffuse downwards decreasing the shielding of CO, which in
turn increases the CO photodissociation rate and thus the surface
density of CI. The higher CI surface density translates to a higher
optical depth that lowers the ionisation rate of carbon and thus the
ionisation fraction. However, as $\alpha_{\rm v}$ increases further
(dashed to dotted line), the diffusion timescale is short enough for
ionised carbon to diffuse downwards, increasing the ionisation
fraction at the midplane despite the much lower ionisation rate.

This variation of the ionisation fraction with height above the disc
midplane is reminiscent of the ionisation state of protoplanetary
discs. In vast regions of protoplanetary discs cosmic rays and stellar
X-rays ionise only the disc upper layers, but do not reach the disc
midplane, leading to MRI-dead zones \citep{Gammie1996, Igea1999}. If
the midplane regions of debris discs are similarly MRI-dead, any
vertical mixing or radial diffusion driven by the MRI would be
considerably weaker than in a well-ionised disc. However, the low
midplane ionisation fractions (as low as ${\sim} 10^{-8}$) we find here
may still be well above the critical value required to make the disc
unstable to the MRI. In protoplanetary discs, that critical value can
be of the order of only $10^{-12}$ \citep[e.g.][]{Gammie1996}. The
stability of weakly ionised debris discs against the MRI remains to be
explored, taking into account the chemical composition that differs
largely from protoplanetary discs and to a lesser degree from well
ionised debris discs considered by \citet{Kral2016b}.

\subsection{Dust scale height}

Whilst in this paper we have focused only on the gas evolution, the
dynamics and evolution of dust can be strongly affected by the
presence of gas \citep{Takeuchi2001, Thebault2005, Krivov2009,
  Lyra2013photoelectric, Pearce2020}. In particular,
\cite{Olofsson2022} recently showed how gas can cause dust grains to
settle towards the midplane if their inclinations are damped by the
gas in a timescale (the stopping time) shorter than their collisional
lifetime. This effect is strongest for the smallest grains in contrast
to what happens in protoplanetary discs where it is the largest grains
that readily settle towards the midplane. In debris discs the gas
densities are low enough such that all grains larger than the blow-out
size have Stokes numbers larger than unity \citep[see Figure 11
  in][]{Marino2020gas}, therefore all bound grains are decoupled from
the gas. However, only the smallest grains have stopping times shorter
than their collisional lifetime. Therefore, if gas densities are high
enough we expect small grains to be more concentrated towards the
midplane than large grains and planetesimals.

One important effect that was neglected by \cite{Olofsson2022} is the
stirring by turbulent diffusion that could reduce the amount
  of inclination damping \citep{Youdin2007stirring}. In the presence
of turbulent diffusion (as considered in this paper), the scale height
of dust is expected to be \citep[Equation 24
  in][]{Youdin2007stirring}\footnote{Note that we have taken the
dimensionless eddy time as 1, which is consistent with MRI simulations
\citep{Fromang2006, Zhu2015}}
\begin{equation}
  H_{\rm d} = H \sqrt{\frac{\alpha_{\rm v}}{\rm St}}\sqrt{\frac{\rm 1+St}{\rm 1+2St}},
\end{equation}
where St is the Stokes number or the dimensionless stopping time. The
Stokes number is defined as $\frac{\upi}{2}\frac{a\rho_{\rm
    s}}{\Sigma}$, where $a$, $\rho_{\rm s}$ and $\Sigma$ are the grain
size, the internal density of grains and the gas surface density,
respectively. Since we expect $\mathrm{St}\gg1$, we have $H_{\rm
  d}\approx H \sqrt{\frac{\alpha_{\rm v}}{\rm 2St}}$. We can evaluate
this for a gas rich debris disc with a surface density of
$10^{-5}\ M_{\oplus}$~au$^{-2}$ (i.e. $M_{\rm gas}{\sim}1\ M_{\oplus}$)
to obtain
\begin{equation}
  \frac{H_{\rm d}}{H} = 0.1 \left(\frac{a}{\rm 1\ \mu m}\right)^{-1/2}\left(\frac{\alpha_{\rm v}}{10^{-2}}\right)^{1/2}\left(\frac{\Sigma}{10^{-5}\ M_{\oplus}\ \mathrm{au}^{-2}}\right)^{1/2}\left(\frac{\rho_{\rm s}}{\rm 1\ g\ cm^{-3}}\right)^{-1/2}.
\end{equation}
This means that although gas drag will damp the inclinations of
$\mu$m-sized grains, these will not completely settle towards the
midplane if the gas surface densities are high
($\gtrsim10^{-5}$~$M_{\oplus}$~au$^{-2}$) and turbulent diffusion is
strong ($\alpha_{\rm v}\gtrsim10^{-2}$). Note that these equations are
only valid for those dust grains with a collisional lifetime much
longer than the stopping time. Therefore, observational estimates of
the scale height of gas and small dust grains could help to constrain
the strength of vertical diffusion and gas models.

\subsection{Radial evolution}

\begin{figure*}
  \centering \includegraphics[trim=0.0cm 0.0cm 0.0cm 0.0cm, clip=true,
    width=0.99\textwidth]{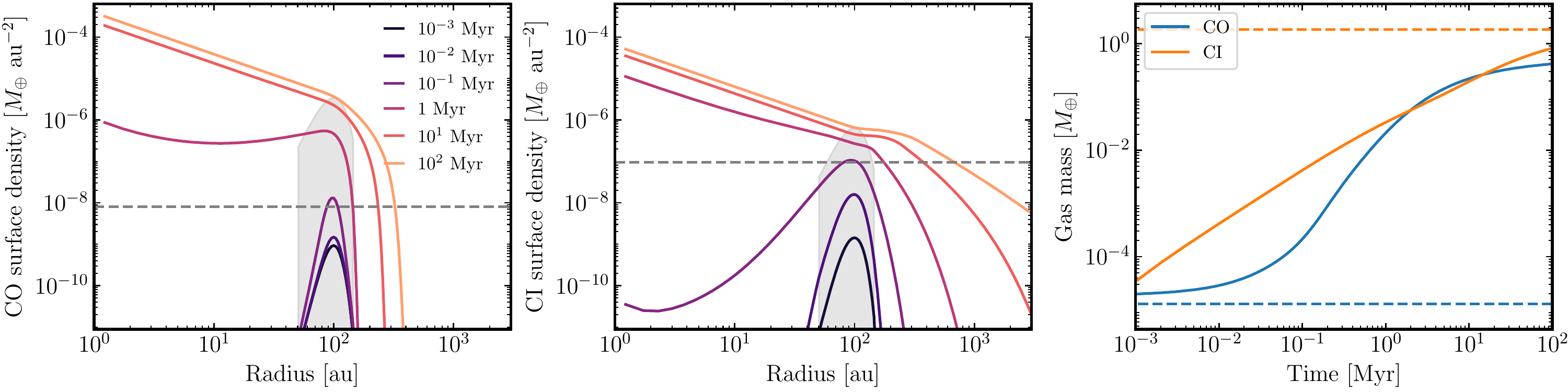}
  \centering \includegraphics[trim=0.0cm 0.0cm 0.0cm 0.0cm, clip=true,
    width=0.99\textwidth]{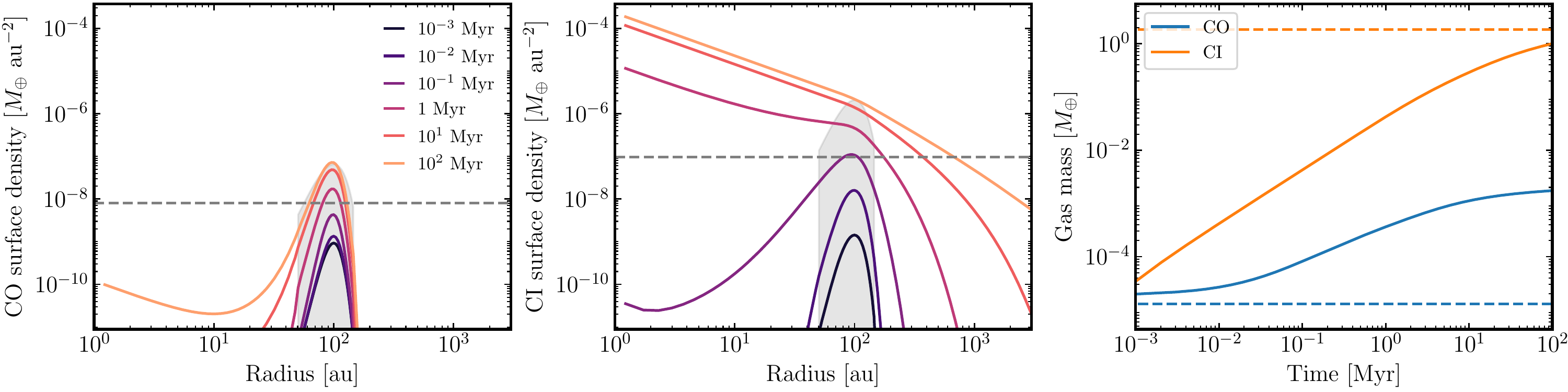}
  \caption{Radial evolution of the surface density of CO (left), CI
    (middle), and their masses (right), in a system where CO gas is
    released at a rate of 0.1 $M_\oplus$~Myr$^{-1}$ from a belt
    centred at 100~au and 50 au wide (FWHM) (grey shaded region), with
    $\alpha=10^{-2}$. The top panel represents a case where CI is in
    the surface of the disc, while in the bottom panel CO and CI are
    perfectly mixed. The horizontal grey dashed lines represent the
    surface density at which CO becomes shielded by a factor $e$. The
    horizontal blue and orange lines in the right panels show the
    expected steady-state mass if CO was unshielded.}
  \label{fig:radial_evolution}
\end{figure*}

Throughout this paper we have shown the importance of the vertical
evolution on the effectiveness of shielding of CO by atomic
carbon. Whether carbon is well mixed or in a layer has profound
implications for how the gas will evolve. Here, we investigate how the
radial evolution could be affected by the vertical diffusion of
gas. To this end, we focus on two extreme scenarios: one where CI is
in a layer completely far above and below the CO-rich midplane, and
another where CO and CI are well mixed. For both scenarios we consider
a system where CO gas is released at a rate of 0.1
$M_\oplus$~Myr$^{-1}$ in a belt centred at 100~au and 50 au wide
(FWHM), around a 1.5~$M_{\odot}$ star (10 $L_{\odot}$), and with
$\alpha=10^{-2}$.

We solve the radial evolution of CO and CI in these two scenarios
using the radial module of the python package \textsc{exogas} that is
based on the model developed in \cite{Marino2020gas}. Because the
photodissociation of CO strongly depends on the disc vertical
structure, which this module does not solve, we need to precompute the
CO photodissociation timescales. We do this using the vertical
evolution module for a grid of 50x50 different surface densities of CO
and CI that are logarithmically spaced. Here is where the two
scenarios differ. In order to compute these timescales, we need to
define the vertical distribution of CO and CI.  First, we assume all
CI is distributed in two surface layers on top of and below the CO
gas. Second, we assume CI and CO are well mixed. In Figure
\ref{fig:tCO_ci} we showed how the CO lifetime could be very different
in these two scenarios. \textsc{exogas} then uses these precomputed
grids and interpolates its values to estimate the CO photodissociation
timescale. These grids provide realistic photodissociation
timescales similar to the ones obtained through photon counting
\citep{Cataldi2020}, but that also account for UV radiation entering
the disc at different angles which lengthens the CO lifetime by a
factor ${\sim}2$ (see \S\ref{sec:photodissociation}).

Figure~\ref{fig:radial_evolution} shows the surface density profiles
of CO (left), CI (middle), and the mass evolution of CO and CI as a
function of time. The top panels show the case where CI is in a
surface layer, and thus the shielding is most effective. Once CI
reaches a high enough surface density
($10^{-7}$~$M_{\oplus}$~au$^{-2}$), CO starts to become shielded. With
a longer lifetime, CO viscously evolves spreading inwards and outwards
beyond the planetesimal belt where CO is released (grey shaded
region). The resulting surface densities at the belt center after
10~Myr of evolution are very similar to those obtained when simulating
the vertical evolution without vertical diffusion
(\S\ref{sec:without_diffusion}). Similar to the simulation results in
\cite{Marino2020gas}, we find a plateau in the CI surface density from
100 to 200~au after 100~Myr of evolution. This feature can be
understood by examining the top left and middle panels. In the first 1
Myr of evolution, CI accumulates forming an accretion disc that is
massive enough to shield CO within 150~au, allowing it to accumulate,
viscously accrete, and dominate the gas surface densities. Beyond the
belt centre, the released CO gas slowly flows outwards, but at a
timescale much longer than the CO lifetime (which decreases
exponentially with radius), resulting in an exponential drop in the
surface density of CO. In this drop between $100-300$~au the gas
composition changes from being CO dominated to being carbon and oxygen
dominated. It is the balance between the decrease in the overall gas
surface density and the increase in the abundance of CI that creates
the plateau. Therefore the plateau starts at the belt centre (where
gas starts to flows outwards) and stops where CI and OI dominate the
bulk of the gas density.

If CI and CO are well mixed (bottom panel) the average CO
lifetime at the belt centre is only 0.02~Myr, which is too short
compared to the viscous timescale (6 Myr). This is why CO does not
viscously spread significantly after 100~Myr. This scenario is the
most problematic for explaining gas-rich discs since their estimated
CI levels would not be enough to shield CO \citep{Cataldi2020}. We
note one caveat in this calculation that results in an underestimation
in the level of shielding compared to the case with a strong diffusion
($\alpha_{\rm v}=\alpha$) and a high gas input rate (0.1
$M_\oplus$~Myr$^{-1}$) presented in \S\ref{sec:results}. We assume
that CI and CO are perfectly mixed, however, when simulating the
vertical evolution with $\alpha_{\rm v}=\alpha$, CO is still more
concentrated towards the midplane than CI. This is because the
photodissociation timescale in the top layers is still shorter than
the diffusion timescale, therefore CO becomes depleted there relative
to the midplane. This behaviour is also shown in
Figure~\ref{fig:vertical_distribution} for values of $\Sigma_{\rm
  ss}=10^{-6}$~$M_\oplus$~au$^{-2}$. The carbon-dominated upper layer
can significantly contribute to shielding CO. In order to completely
mix CO and CI, the diffusion timescale should be shorter than the
unshielded photodissociation timescale. At ${\sim}100$~au, that would
only be obtained with $\alpha\gtrsim10^{-1}$. Therefore, perfect
mixing might never happen in debris discs unless they are strongly
turbulent in the vertical direction.

We conclude that the vertical structure of the gas has a large impact
on the radial evolution. The two extreme scenarios explored here lead
to significantly different results, particularly for the CO
distribution. Constraining the vertical structure of the gas is thus
crucial to advance our understanding of the gas evolution in debris
discs.

\subsection{Observability}
\label{sec:observability}
In this section we produce radiative transfer predictions and discuss
how the vertical distribution of CO and CI would translate into real
observations. To this end, we use
\textsc{RADMC-3D}\footnote{http://www.ita.uni-heidelberg.de/~dullemond/software/radmc-3d}
that allows to produce synthetic images of dust and gas emission based
on several parameters and input files such as the dust distribution,
opacity, gas distribution, velocity field, molecular information,
stellar spectrum, etc. We produce these files and run
\textsc{RADMC-3D} calculations using the \textsc{python} package
\textsc{disc2radmc}\footnote{https://github.com/SebaMarino/disc2radmc}
\citep[based on][]{Marino2018hd107}, which allows to define these
input files and run multiple \textsc{RADMC-3D} calculations in a
simple and concise manner.

We focus on and compare the two main scenarios presented in
\S\ref{sec:results}. In these two scenarios, CO gas has been released
at a rate of 0.1 $M_\oplus$~Myr$^{-1}$ for 10~Myr and $\alpha=10^{-2}$
(i.e. it is in steady state). They differ in that in one gas evolves
without vertical diffusion, while in the other there is strong
vertical diffusion ($\alpha_{\rm v}=\alpha=10^{-2}$). Since our
vertical simulations presented above were one dimensional, we need to
extrapolate their density distribution of CO and CI in order to define
their density as a function of $r$ and $z$. For simplicity, we assume
the vertical distribution (relative to $H(r)$) is independent of
radius and the surface density distribution of both species follows a
Gaussian distribution centred at 110~au and with a standard deviation
of 34~au (FWHM of 80~au). We further assume a scale height of 5~au at
100~au, which scales linearly with radius, and a turbulence
corresponding to $\alpha=10^{-2}$ that broadens slightly the line
emission. Note that although the true gas temperature could be 
different (hotter or colder) and the mean molecular weight higher (if
CO dominated), the vertical distribution relative to $H$ will remain
roughly the same as shown in \S\ref{sec:highT}. This means that while
the vertical extent of CO and CI is uncertain and could vary as
$\sqrt{T/\mu}$, their predicted morphologies presented below are
robust. A higher (lower) temperature would only stretch (compress)
their vertical distributions. Finally, we assume a 1.7~$M_{\odot}$
central star, at a distance of 133~pc, and that is viewed edge-on. The
radial distribution, stellar mass, and distance approximate the values
for HD~32297, a gas-rich edge-on disc \citep{Greaves2016,
  MacGregor2018, Cataldi2020}.

Under these assumptions and density distributions, we compute
synthetic images of CO J$=2-1$ ($\lambda=1300$ $\mu$m) and CI
$^{3}\mathrm{P}_1-^{3}\mathrm{P}_0$ emission ($\lambda=609$ $\mu$m),
both readily accessible through ALMA observations. When calculating
the emission from these species, RADMC-3D assumes Local Thermodynamic
Equilibrium (LTE), i.e. the population levels of each gas species are
set by the local temperature. It is well-known that non-LTE can be
important when gas densities are low \citep[][]{Matra2015}, but the
densities considered here should be high enough for the different
energy levels to be collisionally excited and in thermal equilibrium.

\begin{figure}
  \centering \includegraphics[trim=0.0cm 0.2cm 0.3cm 0.3cm, clip=true,
    width=1.0\columnwidth]{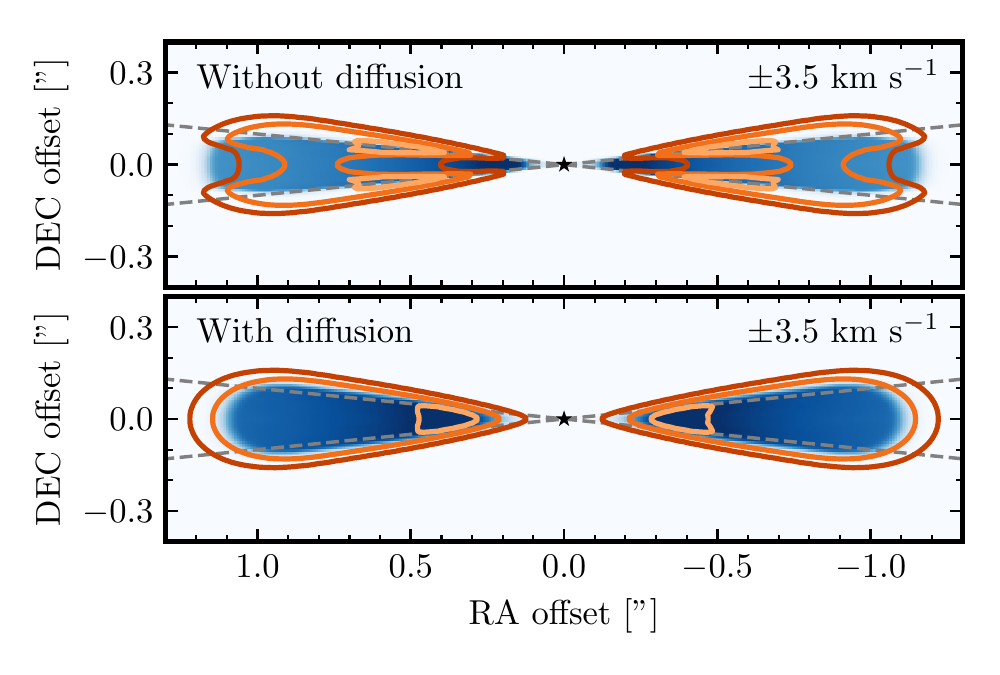}
  \centering \includegraphics[trim=0.0cm 0.2cm 0.3cm 0.3cm, clip=true,
    width=1.0\columnwidth]{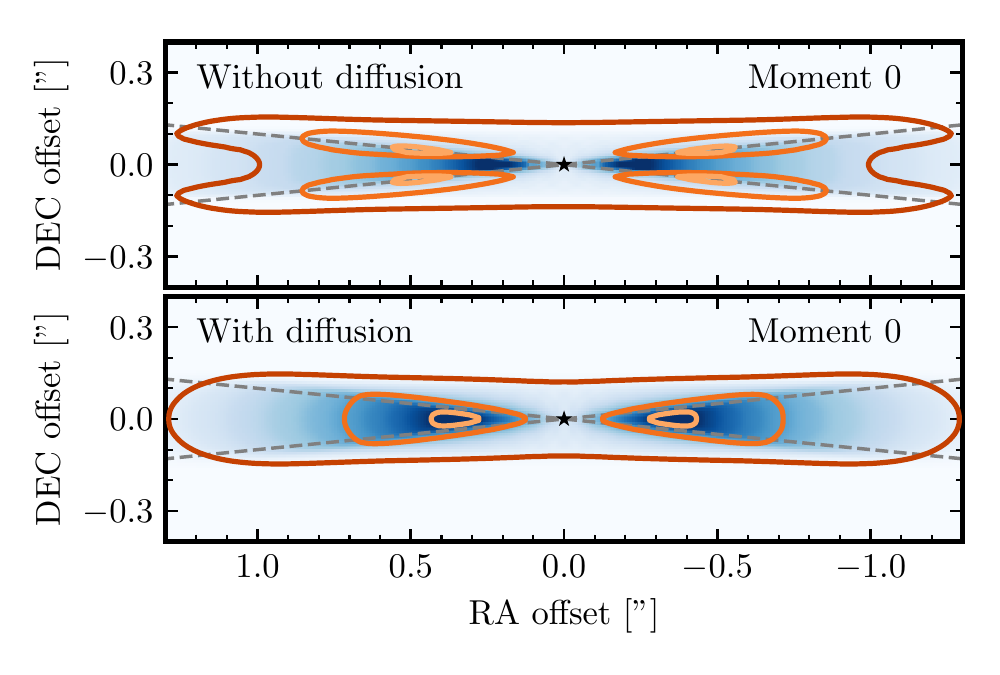}
  \caption{Radiative transfer prediction of CO and CI emission for an
    edge-on disc. The top two panels show the emission at
    line-of-sight velocities of $\pm3.5$~km~s$^{-1}$, while the bottom
    two show the moment 0 (velocity integrated emission). CO J$=2-1$
    emission is shown in blue colours, while CI
    $^{3}\mathrm{P}_1-^{3}\mathrm{P}_0$ emission is shown in orange
    contours at 10, 50 and 90\% of the peak value. The first/third and
    second/fourth panels show a case without and with vertical
    diffusion (respectively), assuming $\alpha=10^{-2}$ and
    $\dot{M}^{+}=10^{-1}$~$M_{\oplus}$~Myr$^{-1}$ after 10~Myr (in
    steady state). The grey dashed line represent 2 scale heights from
    the midplane. }
  \label{fig:prediction}
\end{figure}

Figure \ref{fig:prediction} shows the expected emission at
line-of-sight velocities of $\pm 3.5$~km~s$^{-1}$ (close to the
Keplerian velocity at the belt centre of 3.7~km~s$^{-1}$) and the
moment 0 (i.e. the velocity integrated emission) for a case without
and with diffusion (scenarios presented in Figures
\ref{fig:highmdot_nodiff} and \ref{fig:highmdot_diff},
respectively). The CO intensity is displayed with a blue colour map,
while CI is shown with orange contours. In both the channel maps and
moment 0's we find a similar morphology. Without diffusion, the CO
emission peaks at the midplane and is located within 2 scale heights
from the midplane (grey dashed lines), while CI emission is almost
absent within one scale height and peaks near the upper and lower
edges of the CO emission. Note that the height at which this happens,
relative to the scale height, can vary depending on the gas input rate
and viscosity (\S\ref{sec:vert}). Since CO is very optically thick,
its emission peaks towards the inner regions where the gas is
hotter. Conversely, with diffusion the CI and CO emissions peak in the
midplane displaying a very similar intensity distribution. The only
difference is that CI emission extends vertically slightly further
compared to CO since the latter has a slightly narrower vertical
distribution due to photodissociation. Finally, the observed
  vertical morphology is not sensitive to the radial distribution of
  CO and CI. A different radial distribution would change how the
  emission is distributed radially, but the vertical morphology would
  still be the same. This is especially true for individual channel
  maps where each pixel traces approximately a single radius.

These radiative transfer simulations show that it is possible to
distinguish these two scenarios by looking at the vertical
distribution of CO and CI. Both emission lines have been already
detected and resolved radially in multiple shielded discs. So far,
only \cite{Hughes2017} and \cite{Higuchi2019} have tried to constrain
the vertical extent of CO and CI. Doing a similar analysis for both
lines, \cite{Higuchi2019} found their heights to be consistent with
each other, although with large uncertainties that prevent assessing
the strength of vertical diffusion. Future observations at
high-resolution should reveal if CI is located in a surface layer and
thus indicate the strength of vertical diffusion. Note that with
images that do not resolve the vertical distribution of CO and CI, it
is impossible to distinguish between these two scenarios due to the
unknown gas temperatures and gas scale height.

Whilst in the radiative transfer predictions presented above we focus
in an edge-on case, vertical information can also be obtained for
discs with moderate inclinations. It has been shown that the
characteristic height of emission lines can be obtained from
line-of-sight velocity maps, as well as from individual channels
\citep[e.g.][]{Teague2018, Pinte2018, Casassus2019doppflip,
  Izquierdo2021}. Therefore, ALMA observations at a resolution
comparable or better than the disc scale height offer a unique chance
to determine how strong vertical diffusion is, and thus how important
shielding by CI is.

\subsection{Is the observed gas scale height a good indicator of gas origin?}

It has been suggested that a powerful way to determine the origin of
gas in debris discs is to measure its scale height
\citep[e.g.][]{Hughes2017, Kral2019, Smirnov2021}. By knowing the
scale height and the temperature of a disc, one could estimate the
mean molecular weight. This would allow to determine if the gas is
dominated by molecular Hydrogen or by heavier species such as C, O, or
CO \citep[e.g.][]{Hughes2017}. In addition to the difficulty of
estimating the gas kinetic temperature, a key problem lies in
measuring the gas scale height. To measure it, one would need to use a
gas tracer such as CO or CI, but as shown here the vertical
distribution of these species can differ significantly from the gas
true vertical distribution. For example, in the results presented in
\S\ref{sec:vert} we found that the distribution of CO can be skewed
towards the midplane due to its higher shielding at the midplane if
vertical mixing is weak, mimicking a small scale height. A similar
phenomenon could occur in a primordial gas scenario as well if
vertical diffusion is weak since the CO lifetime will also be longer
near the midplane. This means that the small CO scale height measured
in 49~Ceti by \cite{Hughes2017} might be very different from the true
gas scale height (even if primordial). Hence one should be careful
interpreting a small CO scale height as evidence of a high mean
molecular weight (and thus gas of secondary origin). On the other
hand, the vertical distribution of CI can be much wider mimicking a
larger scale height. Moreover, a high optical depth would rise the
emitting surface of optically thick species making the apparent
vertical distribution wider than in reality. Therefore, we recommend
that scale height values derived directly from observations should be
used with caution, and ideally interpreted using gas models and
radiative transfer simulations. Otherwise, estimates could suffer from
great systematic errors.

\subsection{How does temperature affects the evolution?}
\label{sec:highT}

Throughout the paper we have assumed a gas temperature equal to the
blackbody equilibrium temperature at 100~au of 50~K, which translates
to a scale height of 5~au. However, the gas temperature could be
significantly different from the blackbody and dust temperature in
reality, depending on the gas heating and cooling rates
\citep[e.g.][]{Kral2016}. Therefore, in this section we examine the
evolution of the gas with a temperature 10 times higher (500~K) in
order to understand its effects. The increased temperature changes the
sound speed, which in turn changes other important quantities. First,
the gas scale height is increased by a factor $\sqrt{10}$ from
approximately 5~au to 15~au. Second, the viscosity is 10 times higher
(as we keep $\alpha$ fixed), which translates to a viscous timescale
10 times shorter.

\begin{figure*}
  \centering \includegraphics[trim=0.0cm 0.0cm 0.0cm 0.0cm, clip=true,
    width=0.9\textwidth]{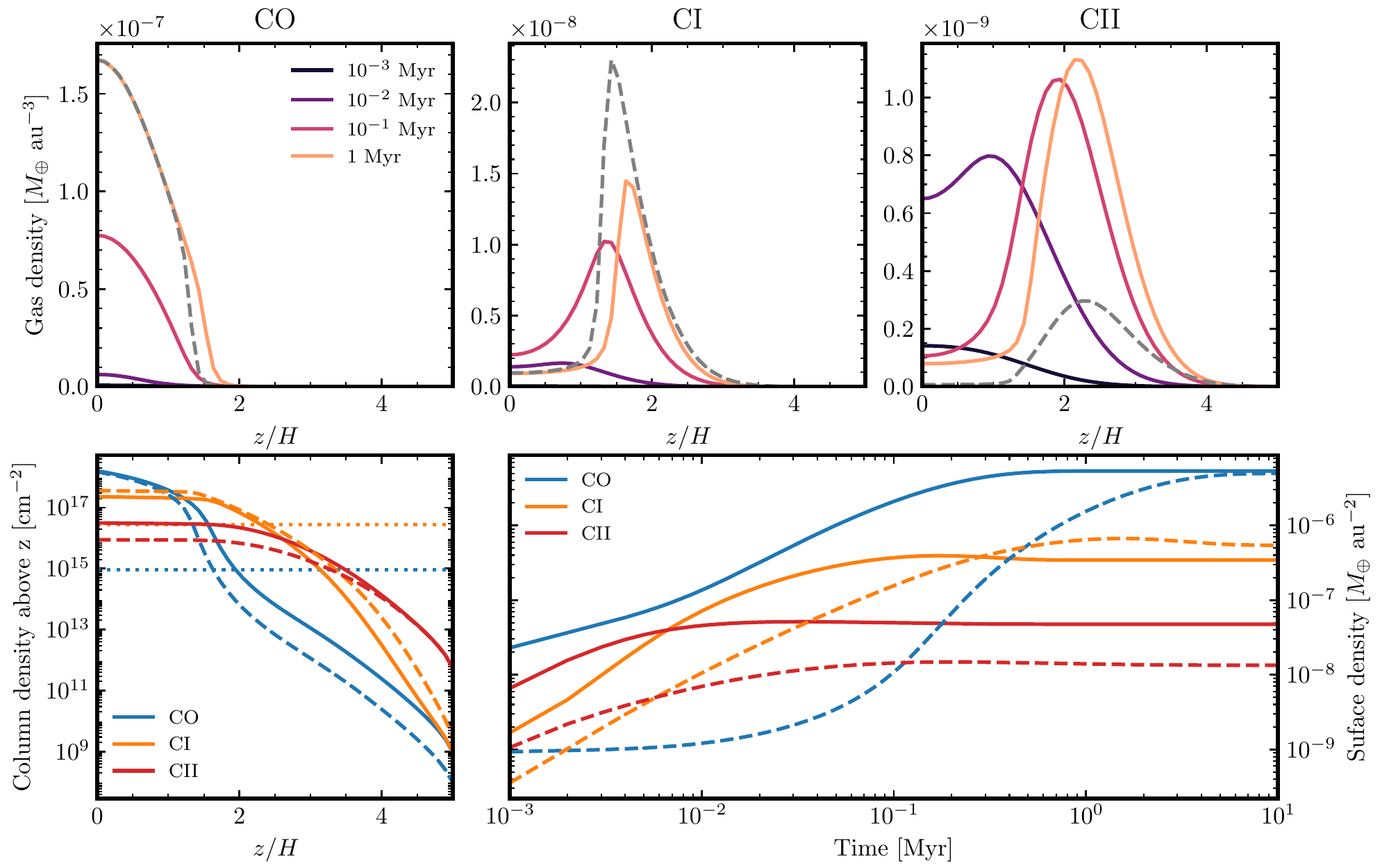}
  \caption{Same as Figure~\ref{fig:lowmdot_nodiff}, but for $T=500$~K
    instead of 50~K, $\dot{M}_{+}=1$~$M_{\oplus}$~Myr$^{-1}$,
    $\alpha=10^{-2}$ and no vertical diffusion. The grey dashed lines
    in the top panels show the steady state gas density when $T=50$~K
    and scaled by a factor $1/\sqrt{10}$ for comparison. Similarly,
    the dashed coloured lines in the bottom left and right panels show
    the steady state column density profiles and the surface density
    evolution for $T=50$~K. }
  \label{fig:highT}
\end{figure*}

We now compare the gas evolution in the case of no vertical diffusion
and a high gas input rate, as presented in
\S\ref{sec:without_diffusion} and in
Figure~\ref{fig:highmdot_nodiff}. In order to focus on the effect of
the temperature alone, we increase the mass input rate by a factor 10
to keep the total gas surface density in steady state
constant. Figure~\ref{fig:highT} shows the evolution of CO, CI and CII
for $T=500$~K. The dashed lines show our standard model with $T=50$~K
for comparison. The grey dashed lines in the top panels have been
scaled by a factor $1/\sqrt{10}$ to account for the difference in
scale height and ease the comparison. We find that the gas evolution
is very similar to the case with a lower temperature. The main
differences are that the CI surface density is ${\sim}30$\% smaller,
the CII surface density is a factor 5 greater, the CO, CI and CII
volumetric densities are lower (due to the increased scale height),
and that the steady state is reached earlier due to the 10 times
shorter viscous timescale. The increase in the surface density of CII
(and decrease of CI) can be explained by the lower volumetric
densities and higher temperature that decrease the recombination rate
of CII. Conversely, reducing the temperature (or increasing $\mu$)
would have the opposite effect: CII recombination rates would be
higher, increasing (reducing) the surface density of CI (CII). These
findings confirm that the evolution of CO and CI is more sensitive to
the column (or surface) densities rather than the volumetric
densities.

If we considered vertical diffusion, the increase in the temperature
would leave the vertical diffusion timescale unchanged
($1/(\alpha_{\rm v}\Omega_{\rm K})$). However, the vertical diffusion
timescale relative to the viscous timescale would become longer, and
thus the CO and CI would become less mixed. We note that this
conclusion is only valid if $\alpha$ and $\alpha_{\rm v}$ are kept
fixed. The high uncertainty in the relative values of $\alpha_{\rm v}$
and $\alpha$, and their dependence on the temperature makes it
impossible to know how different the evolution of a colder vs a hotter
disc will be. Nevertheless, our main conclusion that CO and CI will
become well mixed if $\alpha_{\rm v}> \alpha (H/r)^2$ still holds.

\subsection{Limitations}

The model presented here provides a useful tool to estimate and
quantify the importance of vertical diffusion on the evolution of CO
in exocometary discs. This model, nevertheless, has some limitations
which we discuss here.

First, our model assumes that CO gas is released with a vertical
profile that matches the hydrostatic equilibrium. In reality, the
release of gas will depend on the vertical distribution of solids that
dominate the gas release. If CO gas was released with a high vertical
dispersion, the upper layers would be enriched in CO compared to our
model. Therefore, shielding by carbon would be less
effective. Conversely, if CO was released with a small vertical
dispersion, it would be better shielded by CI. Moreover, we assumed a
constant temperature. The temperature could vary as a function of
height \citep[reaching hundreds or thousands of K at the disc
  surface,][]{Kral2019} or time as the gas densities and composition
change \citep[][]{Kral2016}. Both effects could have an impact on the
level of mixing and thus on the gas evolution if they changed the
ratio of the vertical diffusion and viscous timescales as discussed in
\S\ref{sec:highT}. Exploring these scenarios would require solving the
Navier-Stokes equations together with radiative transfer
equations. Such demanding computations go beyond the scope of this
paper. Additionally, the uncertainty would remain over how $\alpha$
and $\alpha_{\rm v}$ vary as a function of the temperature, density
and ionization fraction (\S\ref{sec:ion}).

Second, our model is one dimensional and thus we cannot track the
radial evolution of the gas. To approximate the viscous evolution, we
remove gas on a viscous timescale at all heights. This might not
necessarily be the case if the viscosity is height
dependant. Nevertheless, this approximation provides a simple way to
remove gas on a certain timescale. Moreover, our model is only
strictly valid at the center of the belt, where CO gas is being
released and there is negligible inflow of gas from another region. To
model the gas interior or exterior to the belt, we would need to know
the composition with which it arrives at that radius. Future modelling
could solve self-consistently the radial and vertical evolution,
allowing to study in more detail this evolution.








\section{Conclusions}
\label{sec:conclusions}

In this paper we have studied for the first time the vertical
evolution of gas in debris discs, focusing on how the vertical
distribution of carbon strongly affects the photodissociating
radiation that CO receives. To this end, we developed a new 1D model
and python package
(\href{https://github.com/SebaMarino/exogas}{\textsc{exogas}}) that
contains a module to simulate the vertical evolution of gas (in
addition to a module to simulate the radial evolution as in
\citealt{Marino2020gas}). Our model takes into account the release of
CO gas from solids in a planetesimal belt, the photodissociation of
CO, the ionisation/recombination of carbon, the radial viscous
evolution, and the vertical mixing due to turbulent diffusion. Below
we summarise our main findings.


First, we determined the efficiency of shielding by CI in two extreme
hypothetical scenarios and confirmed previous findings. If carbon is
primarily located in the surface layer below and above the CO gas, the
lifetime of CO increases exponentially with the column density of CI
\citep[as argued by][]{Kral2019}. However, if CI and CO are vertically
well mixed, shielding by CI is inefficient \citep{Cataldi2020} and the
CO lifetime increases only linearly with the CI column density. This
demonstrates the importance of knowing the true vertical distribution
of gas in debris discs to understand its evolution.




In order to understand what the true vertical distribution should be,
we developed a 1D model to simulate the gas evolution. We showed for
the first time how ISRF photodissociating radiation creates a surface
layer dominated by atomic carbon and oxygen if gas densities are high
and vertical diffusion is negligible. This surface layer acts as an
efficient shield, where CI atoms absorb most of the interstellar UV
photons and become ionised. This process creates three layers where
carbon is in different forms: a top layer dominated by CII where the
recombination rate is slow compared to ionisation, a surface or middle
layer dominated by CI where recombination is efficient and where most
UV radiation is absorbed, and a bottom layer that extends to the
midplane and is dominated by CO. This layered structure allows the
build up of a CO-rich midplane. However, if diffusion is strong CI and
CO become well mixed, exposing CO to a stronger UV radiation. In this
case, shielding by CI is inefficient and thus higher gas release rates
would be required to achieve the same CO gas surface densities than
with a weak vertical diffusion.

Since the strength of the turbulent diffusion ($\alpha_{\rm v}$) is
uncertain in debris discs, we explored a wide range of values to
characterise the different behaviours. We found that in steady state
the vertical structure of the gas is mainly set by the vertical
diffusion timescale ($H^2/D$) relative to the viscous timescale
(i.e. the gas removal/replenishment timescale, $r^2/\nu$). In other
words, if $\alpha_{\rm v}/\alpha > (H/r)^2$ diffusion happens faster
than the gas replenishment and thus CO and CI are well
mixed. Conversely, if $\alpha_{\rm v}/\alpha < (H/r)^2$ diffusion
plays a minor role and a surface layer forms that is dominated by CI
and oxygen, which shields the CO underneath. If the vertical diffusion
timescale is longer than the age of the system, the CI gas could still
form a surface layer if it is optically thick. Note that if gas
densities are low enough such that the gas is optically thin to UV
radiation in the vertical direction, CI and CO have very similar
distributions independently of the diffusion strength.

Based on our results, we discussed what these findings imply for the
known gas-rich debris discs. If diffusion is weak, the simple
secondary scenario could still explain the population of gas-rich
debris discs and account for the estimated CI masses. However, if
observations of gas-rich debris discs indicate that CI and CO are well
mixed, the standard secondary origin scenario might need
reevaluation. Unfortunately, the strength of vertical diffusion in
debris discs is unknown. If turbulence is driven by the MRI,
$\alpha_v$ might be similar to $\alpha$ which would result in a
well-mixed CO and CI gas. However, this picture could be even more
complicated since we find that the ionisation varies greatly with
height, which could make any MRI-driven diffusion height dependent.

We also discussed how turbulent diffusion could affect the vertical
distribution of small grains. Those grains are susceptible to gas drag
that damps their inclinations and thus should settle towards the
midplane \citep{Olofsson2022}. However, turbulent diffusion can
counterbalance this effect and limit the amount of
settling. Measurements of the gas and small dust heights could provide
constraints to the strength of vertical diffusion.

Whilst our simulations focused on the gas evolution at the belt
centre, we showed how the efficiency of shielding has an effect on the
viscous radial evolution of CO. Using radial evolution simulations we
showed that if CO and CI are well mixed (and thus shielding by CI is
weak) the CO lifetime might not be long enough for it to spread
interior to the planetesimal belt. Therefore, the vertical evolution
of the gas can have a profound effect on the radial distribution of
CO.


Although vertical diffusion is largely unconstrained, using our
simulations and radiative transfer predictions we showed that
high-resolution ALMA observations could reveal if CO and CI are well
mixed or segregated into different layers. Such observations would
constrain $\alpha_{\rm v}/\alpha$ and ultimately determine how
efficient shielding by CI is. This could confirm the exocometary
origin of the gas, or favour a primordial origin where shielding is
mostly done by hydrogen.


Finally, we discussed how attempts to infer the gas scale height
directly from observations could suffer from large systematic
errors. This is because the vertical distribution of CO and CI can be
significantly different from the total gas distribution in gas-rich
discs. CO gas can be concentrated towards the midplane, while CI can
have an effective scale height much larger than the true value. We
thus recommend the use of multiple species and the comparison with
models as the one presented in this paper.

\section*{Acknowledgements}
S. Marino is supported by a Junior Research Fellowship from Jesus
College, University of Cambridge. G. Cataldi is supported by the NAOJ
ALMA Scientific Research grant code 2019-13B. M. Jankovic is supported
by the UK Science and Technology research Council (STFC) via the
consolidated grant ST/S000623/1.

\section*{Data availability}
The data underlying this article will be shared on reasonable request
to the corresponding author. \textsc{exogas} can be downloaded from
\url{https://github.com/SebaMarino/exogas}, where readers can also
find examples of how to use it. \textsc{disc2radmc} can be downloaded
from \url{https://github.com/SebaMarino/disc2radmc}, where examples are
also available.





\bibliographystyle{mnras}
\bibliography{SM_pformation} 



\appendix

\section{Parameters}

In Table~\ref{tab:parameters} we summarise the most important
parameters that we use.
\begin{table*}
  \centering
  \caption{Model parameters, their default units, and description. }
  \label{tab:parameters}
  \begin{adjustbox}{max width=1.0\textwidth}
    \begin{tabular}{c c l } 
  \hline
  \hline
  Parameter & Unit & Description\\
  \hline
  $r$                          & au & Planetesimal belt radius. \\
  $\Delta r$                   & au & Full-width-half-maximum of planetesimal belt. \\
  $z$                          & au & Height above the disc midplane. \\
  $H$                          & au & Vertical disc scale-height as defined in Eq.~\ref{eq:H}. \\ 
  $\Sigma$                    & $M_{\oplus}$~au$^{-2}$ & Total surface density of gas. \\
  $\rho(z)$                    & $M_{\oplus}$~au$^{-3}$ & Total density of gas at height $z$. \\
  $\rho_{i}(z)$                & $M_{\oplus}$~au$^{-3}$ & Density of species $i$ (CO, CI or CII gas) at height $z$. \\
  $\Omega_{\rm K}$              & rad yr$^{-1}$   & Keplerian frequency. \\
  $c_{\rm s}$                   & m s$^{-1}$  & Isothermal sound speed as defined in Eq.~\ref{eq:cs}. \\
  $\mu$                        &  ...  &  Mean molecular weight. \\
  $m_{\rm p}$                   &  kg & Proton mass. \\
  $L_{\star}$                   & $L_{\odot}$ & Stellar luminosity. \\
  $T$                          & K & Gas temperature as defined in Eq.~\ref{eq:T}. \\
  $\lambda$                    & nm & Wavelength. \\
  $\theta$                     & rad & Polar angle measured from the vertical direction. \\
  $\sigma_{\rm ph}(\lambda)$    & cm$^2$ & CO photodissociation cross section from \cite{Heays2017}. \\
  $\sigma_{ \rm CI}$            & cm$^2$ & CI ionisation cross section from \cite{Heays2017}. \\
  $\sigma_i(\lambda) $         & cm$^{2}$ & cross section of species $i$. \\
  $\phi_{\lambda}$              & s$^{-1}$ cm$^{-2}$ nm$^{-1}$  & Interstellar radiation field from \cite{Draine1978} with extension from \cite{vanDishoeck1982}. \\
  $\tau_i(\lambda, \theta, z)$ & ... & Optical depth of species i in the direction $\theta$ at height $z$ and wavelength $\lambda$  as defined in Eq.~\ref{eq:tau}. \\ 
  $R_{\rm ph, 0}$               & yr$^{-1}$ & Unshielded CO photodissociation rate per molecule as defined in Eq.~\ref{eq:Rph0}.   \\
  $R_{\rm ph}(z)$               & yr$^{-1}$ & CO Photodissociation rate per molecule at height z as defined in Eq.~\ref{eq:Rph}. \\
  $K(N_{\rm CO})$               & ... & Shielding factor due to CO self-shielding as defined in Eq.~\ref{eq:K}. \\
   $\dot{\rho}_{\rm ph}(z)$       & $M_{\oplus}$~au$^{-3}$~yr$^{-1}$ & CO Photodissociation rate per unit volume at height $z$ as defined in Eq.~\ref{eq:rhoph}.  \\
  $R_{\rm ion, 0}$                & yr$^{-1}$ & Ionisation rate per CI atom in the optically thin regime as defined in Eq.~\ref{eq:Rion0}. \\
  $R_{\rm ion}(z)$               & yr$^{-1}$ & Ionisation rate per CI atom as defined in Eq.~\ref{eq:Rion}. \\
  $R_{\rm rc}(z)$                & yr$^{-1}$ & Recombination rate per CII atom as defined in Eq.~\ref{eq:Rrc}. \\
  $\alpha_{\rm rc}(T)$           & au$^3$ yr$^{-1}$ & Recombination rate coefficient. \\
  $n_{\rm e^{-}}(z)$              & au$^{-3}$ & Electron number density. \\
  $\dot{\rho}_{\rm ion}(z)$      & $M_{\oplus}$~au$^{-3}$~yr$^{-1}$ &  Net ionisation rate per unit volume at height $z$ as defined in Eq.~\ref{eq:rhoion}. \\
  $\dot{\rho}^{+}$              & $M_{\oplus}$~au$^{-3}$~yr$^{-1}$ & CO gas release rate per unit volume as defined in Eq.~\ref{eq:rhop}. \\
  $\dot{\Sigma}^{+}$            & $M_{\oplus}$~au$^{-2}$~yr$^{-1}$ & CO gas release rate per unit surface as defined in Eq.~\ref{eq:sigmap}. \\
  $\dot{M}^{+}$                 & $M_{\oplus}$~yr$^{-1}$ & CO gas release rate. \\
  $\dot{\rho}_{\rm vis, i}(z)$   & $M_{\oplus}$~au$^{-3}$~yr$^{-1}$ & Mass loss due to viscous evolution as defined in Eq.~\ref{eq:rhovis} \\
  $t_{\rm vis}$               & yr  & Viscous timescale as defined in Eq.~\ref{eq:tvis}.     \\
  $t_{\rm diff}$               & yr  & Vertical diffusion timescale as defined in Eq.~\ref{eq:tdiff}.     \\

  $\nu$                         & au$^2$ yr$^{-1}$ & Kinematic viscosity defined as $\alpha c_{\rm s}H$. \\ 
  $\alpha$                      & ... & Dimensionless viscosity parameter. \\ 
  $D$                         & au$^2$ yr$^{-1}$ & Diffusion coefficient defined as $\alpha_{\rm v}c_{\rm s}H$. \\ 
  $\alpha_{\rm v}$                      & ... & Dimensionless vertical diffusion parameter.       \\
  $\dot{\rho}_{i, D}$           & $M_{\oplus}$~au$^{-3}$~yr$^{-1}$ & Diffusion term as defined in Eq.~\ref{eq:diff}.  \\
  \hline
  \end{tabular}  
  \end{adjustbox}
\end{table*}

\section{Vertical structure with $\alpha=10^{-1}$}
\label{a:vertical}
\begin{figure}
  \centering \includegraphics[trim=0.0cm 0.0cm 0.0cm 0.0cm, clip=true,
    width=1.0\columnwidth]{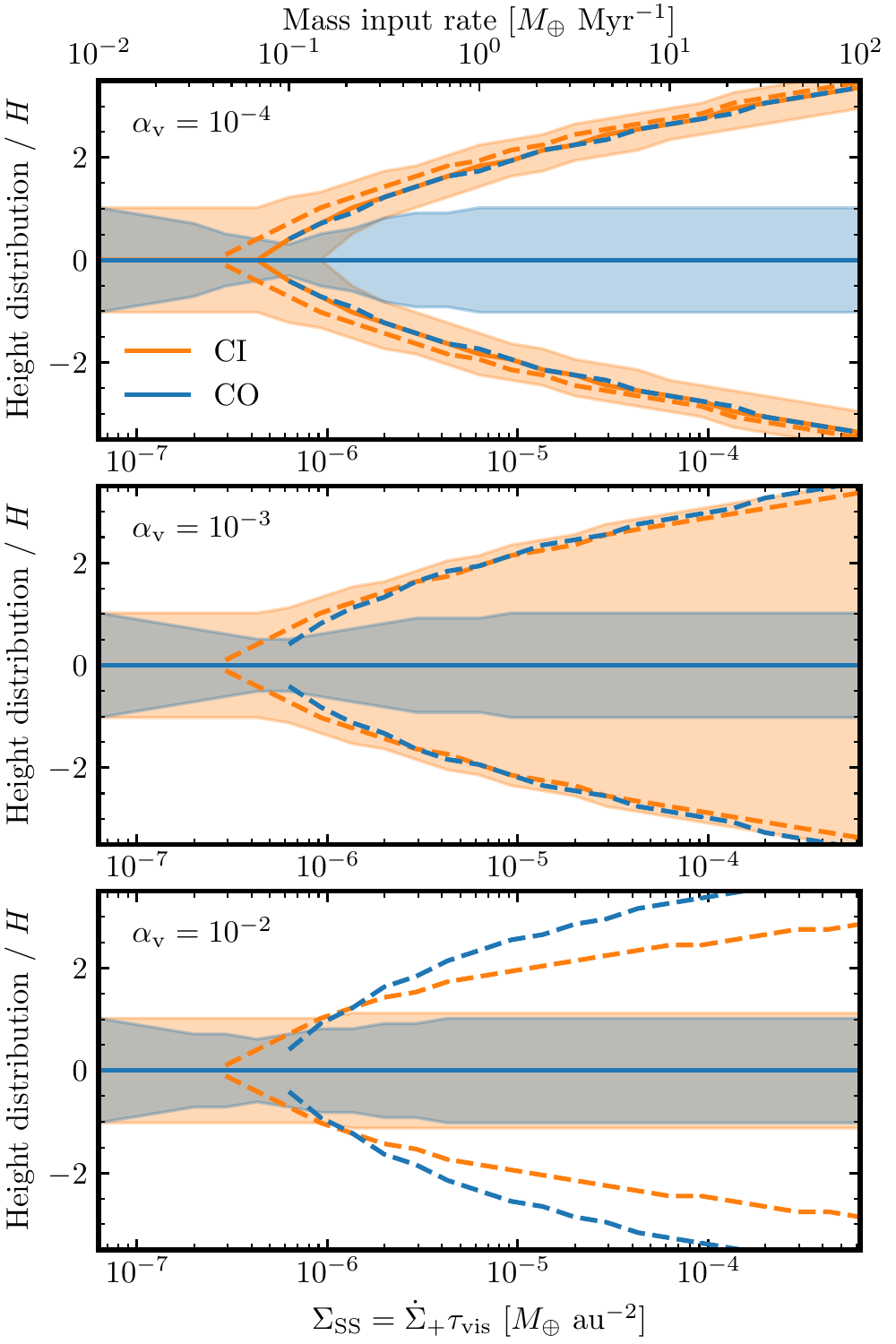}
  \caption{Vertical distribution in steady state of CO (blue) and CI
    (orange) normalized by the scale height ($H$) as a function of the
    steady state surface density of gas, in a case with
    $\alpha=10^{-1}$. The solid line shows where the density
    distribution peaks, while the shaded region represents the region
    where the density is at least 60\% of the peak density ($1\sigma$
    drop). The dashed lines represent the height at which CI and CO
    become optically thick to photodissociating radiation in the
    vertical direction. The top, middle, and bottom panels show
    scenarios where $\alpha_{\rm v}$ is $10^{-4}$, $10^{-3}$, and
    $10^{-2}$, respectively. The upper x-axis in the top panel shows
    the mass input rates as reference.}
  \label{fig:vertical_distribution2}
\end{figure}

In Figure \ref{fig:vertical_distribution2} we show the vertical
distribution of CO and CI as a function of the steady state surface
density of gas, and for three different values of $\alpha_{\rm v}$:
$10^{-4}$, $10^{-3}$, and $10^{-2}$. These correspond to the same
ratios of $\alpha_{\rm v}/\alpha$ than those shown in
\S\ref{sec:vert}. This demonstrates how the vertical distribution of
CO and C will mostly depend on $\alpha_{\rm v}/\alpha$ rather than
their absolute values.

\bsp	
\label{lastpage}
\end{document}